\documentclass[letter]{aa}

\usepackage{graphicx}
\usepackage[font=small]{caption}
\usepackage{indentfirst}
\usepackage{tabu}
\usepackage{ragged2e}
\usepackage[utf8]{inputenc}
\usepackage{soul}

\usepackage{natbib}
\usepackage{booktabs}
\usepackage{fixltx2e}

\usepackage{amssymb}
\usepackage{txfonts}
\usepackage[a]{esvect}
\usepackage{tikz}

\usepackage{url}
\usepackage{hyperref}

\everymath={\rm}

\hypersetup{
   colorlinks=true,       % false: boxed links; true: colored links
   linkcolor=black,       % color of internal links (change box color with linkbordercolor)
   citecolor=blue,        % color of links to bibliography
   urlcolor=black         % color of external links
}

\newcommand{\degree}[0]{$^{\circ}$}
\newcommand{\cbra}[1]{\left( #1 \right)}      % put the argument between parentheses (curve brackets)

\makeatletter
\newcommand{\rmnum}[1]{{\footnotesize{\expandafter\@slowromancap\romannumeral #1@}}}
\newcommand{\Rmnum}[1]{{\expandafter\@slowromancap\romannumeral #1@}}
\makeatother

\title{Physical conditions in Centaurus A's northern filaments \\ \Rmnum{2}. Does the HCO$^+$ emission highlight the presence of shocks?\thanks{This paper makes use of the following ALMA data: ADS/JAO.ALMA$\#$2015.1.01019.S and $\#$2016.1.00261.S.}}

\author{
   Q. Salom\'e$^{1,2}$,
   P. Salom\'e$^{3}$,
   B. Godard$^{3,4}$,
   P. Guillard$^{5}$,
   A. Gusdorf$^{4,3}$
}

\institute{
   Finnish Centre for Astronomy with ESO (FINCA), University of Turku, Vesilinnantie 5, 20014 Turku, Finland \\ email: quentin.salome@utu.fi \and
   Aalto University Mets\"ahovi Radio Observatory, Mets\"ahovintie 114, 02540 Kylm\"al\"a, Finland \and
   Observatoire de Paris, LERMA, CNRS, Universit\'e PSL, Sorbonne Universit\'e, 75014 Paris, France \and
   Laboratoire de Physique de l’ENS, Ecole Normale Sup\'erieure, Universit\'e PSL, CNRS, Sorbonne Universit\'e, 75005 Paris, France \and
   Sorbonne Universit\'e, CNRS, Institut d’Astrophysique de Paris, 98bis bd Arago, 75014 Paris, France
}

\date{Received 31 May 2024 / Accepted 15 September 2024}

\titlerunning{HCO$^+$(1--0) and HCN(1--0) with ALMA in the northern filaments of Centaurus A}
\authorrunning{Salom\'e et al.}

\abstract{
   We present the first observations of HCO$^+$(1--0) and HCN(1--0) emission in the northern filaments of Centaurus A with ALMA. HCO$^+$(1--0) is detected in nine clumps of the Horseshoe complex, with similar velocities as the CO(1--0) emission. Conversely, HCN(1--0) is not detected, and we derive upper limits on the flux.
At a resolution of $\sim 40\: pc$, the line ratio of the velocity-integrated intensities $I_{HCO+}/I_{CO}$ varies between 0.03 and 0.08, while $I_{HCO+}/I_{HCN}$ is higher than unity, with an average lower limit of 1.51. These ratios are significantly higher than what is observed in nearby star-forming galaxies. Moreover, the ratio $I_{HCO+}/I_{CO}$ decreases with increasing CO-integrated intensity, contrary to what is observed in the star-forming galaxies. This indicates that the HCO$^+$ emission is enhanced and may not arise from dense gas within the Horseshoe complex. This hypothesis is strengthened by the average line ratio $I_{HCN}/I_{CO}<0.03$, which suggests that the gas density is rather low.

   Using non-local thermal equilibrium, large velocity gradient modelling with RADEX, we explored two possible phases of the gas, which we call 'diffuse' and 'dense' and are characterised by a significant difference in the HCO$^+$ abundance relative to CO, respectively $N_{HCO+}/N_{CO}=10^{-3}$ and $N_{HCO+}/N_{CO}=3\times 10^{-5}$.
The average CO(1--0) and HCO$^+$(1--0) integrated intensities and the upper limit on HCN(1--0) are compatible with both diffuse ($n_H=10^3\: cm^{-3}$, $T_{kin}=15-165\: K$) and dense gas ($n_H=10^4\: cm^{-3}$, $T_{kin}>65\: K$).
The spectral setup of the present observations also covers SiO(2--1). While undetected, the upper limit on SiO(2--1) is not compatible with the RADEX predictions for the dense gas.
We conclude that the nine molecular clouds detected in HCO$^+$(1--0) are likely dominated by diffuse molecular gas.
While the exact origin of the HCO$^+$(1--0) emission remains to be investigated, it is likely related to the energy injection within the molecular gas that prevents gravitational collapse and star formation.
}


\begin{document}

\maketitle

%%%%%%%%%%%%%%%%%%%%%%%%%%%%%%%%%%%%%%%%%%%%%%%%%%%%%%%%%%%%%%%%%%%%%%%%%%%%%%%%%%%%%%%%%%%%%%%%%%%%%%%%%%%%

\section{Introduction}

   Active galactic nuclei (AGN) are believed to play a role in regulating and/or quenching star formation in galaxies \citep{Bower_2006,Croton_2006}. However, evidence of AGN positive feedback, which enhances star formation, is also seen in radio galaxies, in particular in regions of radio jet-gas interaction at low \citep{Croft_2006, Morganti_2010,SalomeQ_2015a,Zovaro_2020,Capetti_2022} and high redshifts \citep{Nesvadba_2020,Duncan_2023}.

   Centaurus A is a large double-lobed radio source that extends over about 600 kpc (see \citealt{Neff_2015a} and references therein). About 15 kpc away from the radio core, the radio lobe emission encounters a large H\rmnum{1} shell \citep{Schiminovich_1994} associated with recent star formation \citep{Rejkuba_2001,Auld_2012,Joseph_2022}.
Its proximity ($d=3.8\: Mpc$; \citealt{Harris_2010}) makes Centaurus A the ideal target for studying the impact of an AGN-driven jet on its environment.
\cite{SalomeQ_2016a, SalomeQ_2016b} observed a large reservoir of molecular gas at the intersection of the radio continuum and the H\rmnum{1} gas, within the northern optical filaments \citep{Blanco_1975,Morganti_1991}.
The bulk of the molecular gas lies outside the H\rmnum{1} shell, suggesting that the jet-gas interaction induced star formation by triggering the atomic-to-molecular gas phase transition \citep{SalomeQ_2016b}. However, the large reservoir of molecular gas is very inefficient at forming stars compared with nearby star-forming galaxies \citep{SalomeQ_2016a,SalomeQ_2016b}.

   The Atacama Large Millimeter/submillimeter Array (ALMA) resolved the CO emission in the so-called northern filaments of Centaurus A and enabled a collection of giant  molecular clouds (GMCs) distributed along a Horseshoe-like structure to be observed \citep{SalomeQ_2017}. These molecular clouds have very similar physical properties (mass, size, and velocity dispersion) as in the inner Milky Way. However, the virial parameter indicates that kinetic energy is injected into the molecular clouds and prevents gravitational collapse. Moreover, the excitation of the ionised gas associated with the Horseshoe complex suggests the presence of shocks \citep{SalomeQ_2016b,SalomeQ_2017}, which could explain the inefficient jet-induced star formation.

%%%---------------------------------------

\begin{table*}[h]
  \centering
  \small
  \caption{\label{table:overview} ALMA and ACA observations during Cycle 3 and Cycle 4.}
  \begin{tabular}{llcccccc}
    \hline \hline
    Species \& Line & Observatory &    Project     & $\nu_{obs}$ &      Resolution       &   MRS    &  $\delta_v$   &    rms     \\
                    &             &                &    (GHz)    &                       &          & ($km.s^{-1}$) & (mJy/beam) \\ \hline
    CO(1--0)        & ALMA-12m    & 2015.1.01019.S &  115.0611   & $1.48''\times 1.10''$ & $10.9''$ &      1.5      &     4.9    \\
    CO(1--0)        & ALMA+ACA    & 2015.1.01019.S &  115.0611   & $1.56''\times 1.15''$ & $57.5''$ &      1.5      &     4.7    \\
    HCN(1--0)       & ALMA-12m    & 2016.1.00261.S &   88.4689   & $2.30''\times 2.00''$ & $15.5''$ &      3.0      &     0.58   \\
    HCO$^+$(1--0)   & ALMA-12m    & 2016.1.00261.S &   89.0259   & $2.29''\times 1.99''$ & $15.4''$ &      3.0      &     0.59   \\
    SiO(2--1)       & ALMA-12m    & 2016.1.00261.S &   86.6887   & $2.35''\times 2.03''$ & $15.8''$ &      3.0      &     0.60   \\ \hline
  \end{tabular}
\end{table*}

\begin{figure*}[h]
  \centering
  \sidecaption
  \includegraphics[width=0.65\linewidth,trim=90 50 130 100,clip=true]{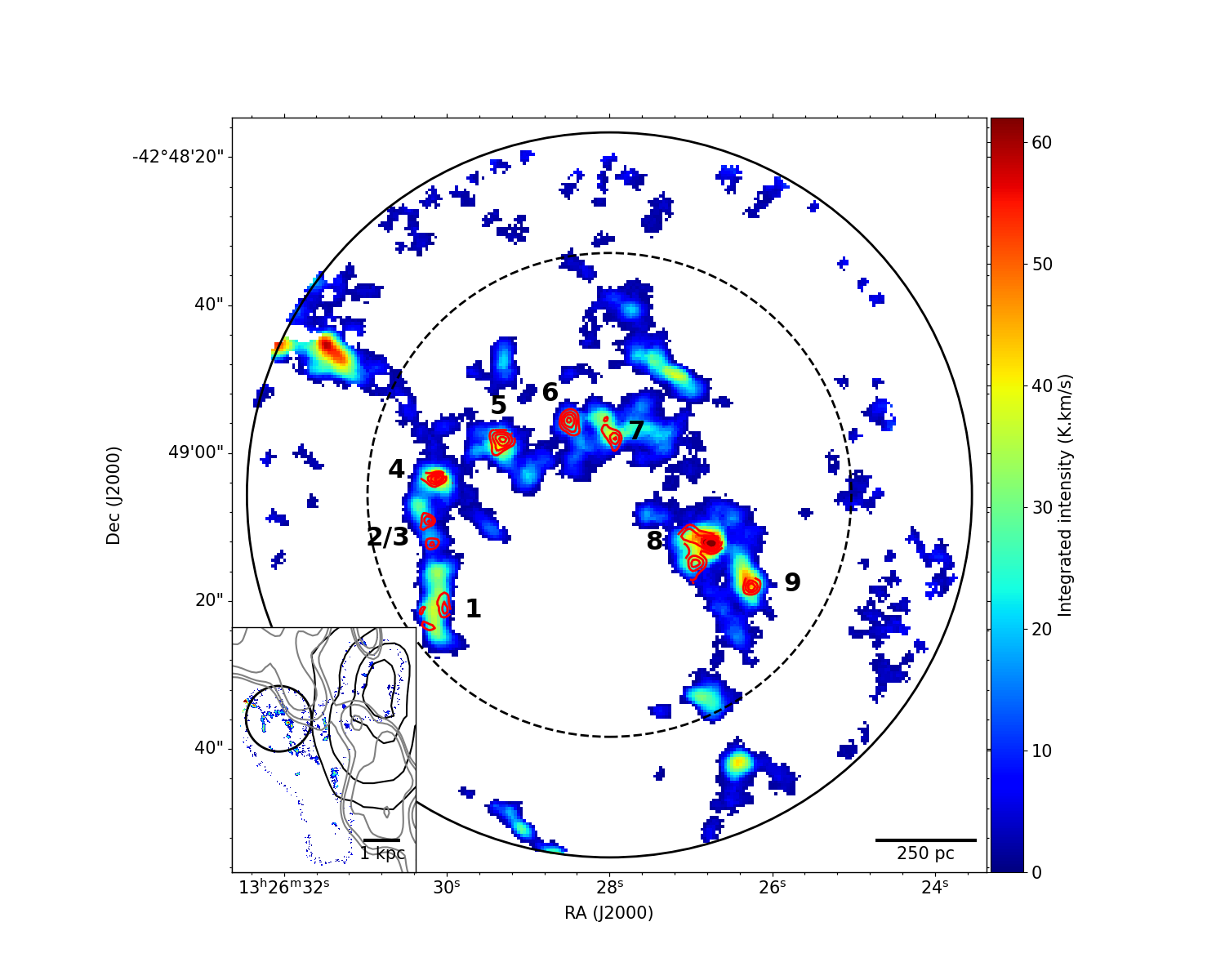}
  \caption{\label{moment0} Moment 0 map of the CO(1--0) emission at the resolution of the HCO$^+$ observations. The red contours are those of the HCO$^+$ emission. The solid line and dashed circles correspond to the field of view and the FWHM of the HCO$^+$ primary beam, respectively. The numbers are the clump labels.
  \emph{Subpanel:} Position of the HCO$^+$ field of view with respect to the H\rmnum{1} shell (black contours) and the radio continuum (grey contours). A larger spatial overview is provided in Fig. 1 of \cite{Oosterloo_2005}, where the region is labelled 'outer filament'.}
\end{figure*}

%%%---------------------------------------

   \cite{SalomeQ_2019} used mid-J CO lines observed with the Atacama Pathfinder Experiment (APEX) in position 16 from \cite{SalomeQ_2016b}. Using the Paris-Durham shock model \citep{Flower_2015}, they show that the CO line ratios are compatible with low-velocity shocks in diffuse gas (shock velocities between 4 and $20\: km.s^{-1}$ and pre-shock density $n_H=100\: cm^{-3}$). However, these predictions are an average over the large area covered by the beams ($>250\: pc$). The high resolution of ALMA now allows the shocks experienced by the different GMCs to be studied.

   In this Letter we present ALMA observations of dense gas tracers (HCN and HCO$^+$) in the Horseshoe complex and compare them with the cold molecular gas traced by CO. We aim to constrain the properties of the molecular gas when the GMCs are resolved. The data are presented in Sect. \ref{sec:Obs}. We analyse the data in Sects. \ref{sec:Res} and \ref{sec:line_ratios}, and then compare our observations with a grid of RADEX models in Sect. \ref{sec:RADEX}.

\section{Observations}
\label{sec:Obs}

   The HCN(1--0) and HCO$^+$(1--0) lines were observed with the ALMA 12m array during Cycle 4 using the Band 3 receivers (project 2016.1.00261.S). The observations consist of one pointing centred on the eastern CO-bright region from \cite{SalomeQ_2016b}, which was later identified as the Horseshoe complex by \cite{SalomeQ_2017}.
The baselines ranged from 15.1~m to 460~m, corresponding to a resolution of $2.1''\times 1.6''$ and maximum recoverable scales (MRS) of about $15.5''\sim 280\: pc$.

   To improve the sensitivity, we decreased the spatial resolution by cutting any baselines larger than 272.6~m (corresponding to the 90th percentile), and we reduced the channel sampling to $3.0\: km.s^{-1}$. The total integration time of 2.1~h provides a noise level of 0.58--0.59~mJy/beam in a synthesised beam of $2.3''\times 2.0''\sim 42\times 37\: pc$ ($PA\sim 88$\degree).
Our observations also covered the SiO(2--1) line with a rms of 0.60~mJy/beam.

   For the CO(1--0) emission, we re-calibrated, re-combined, and re-imaged the observations (\citealt{SalomeQ_2017,SalomeQ_2019}; project 2015.1.01019.S) because an issue with CASA affected the flux in the Atacama Compact Array (ACA) data (see NAASC Memo 117\footnote{\url{https://drive.google.com/file/d/1F2LoWnJlRTdq-hRrlPfHfz78sfdhBRzM/view}}).
We obtained a spatial resolution of $1.56''\times 1.15''\sim 29\times 21\: pc$ ($PA=89$\degree). The noise level is 4.7~mJy/beam at a spectral resolution of $1.5\: km.s^{-1}$.

   Table \ref{table:overview} summarises the main characteristics of the various data cubes. The combination of ALMA and ACA data for the CO(1--0) emission enabled us to cover a large range of spatial scales and limit the spatial filtering by the interferometer.
For HCN, HCO$^+$, and SiO, the maximum recovered scale corresponds to 280--290 pc. The HCN and HCO$^+$ emission is commonly used as a tracer of dense gas in galaxies. Moreover, in local GMCs, most of the emission is localised in the densest regions (e.g. in Orion B; \citealt{Pety_2017}). While we cannot rule out a filtering of possible extended HCN and HCO$^+$ emission, we expect the emission to be much more compact than the MRS, and the spatial filtering to be negligible. We tested this hypothesis by comparing our results with those obtained when considering the CO(1--0) emission from ALMA alone, which has a MRS of $10.9''$ (i.e. $\sim 200\: pc$) closer to the MRS of HCN, HCO$^+$, and SiO.

%%%---------------------------------------

\begin{table*}
  \centering
  \small
  \caption{\label{table:spec} Properties of the different clumps detected with ALMA.}
  \begin{tabular}{lcccccc|ccc}
    \hline \hline
    Clump   & $\int S_{HCO^+}d\varv$ &      FWHM      &   $v_{peak}$   & $\int S_{CO}d\varv$ & $\int S_{HCN}d\varv$  & $\int S_{SiO}d\varv$ & $I_{HCO+}/I_{CO}$ & $I_{HCO+}/I_{HCN}$ & $I_{HCN}/I_{CO}$ \\
            &   ($mJy\,km\,s^{-1}$)  & ($km\,s^{-1}$) & ($km\,s^{-1}$) &  ($Jy\,km\,s^{-1}$) &  ($mJy\,km\,s^{-1}$)  & ($mJy\,km\,s^{-1}$)  &                   &                    &                  \\ \hline
    1       &      $41.8\pm 7.2$     &   $37\pm 7$    &  $\sim -206$   &    $1.23\pm 0.04$   &       $<32.3$         &       $<37.9$        &  $0.057\pm 0.012$ &       $>1.28$      &     $<0.044$     \\ % 29 pix
    2       &      $20.0\pm 3.3$     &   $28\pm 5$    &  $\sim -237$   &    $0.43\pm 0.03$   &       $<15.7$         &       $<15.9$        &  $0.078\pm 0.018$ &       $>1.26$      &     $<0.062$     \\ % 14 pix
    3       &      $22.9\pm 4.8$     &   $24\pm 8$    &  $\sim -239$   &    $0.53\pm 0.03$   &       $<17.0$         &       $<16.2$        &  $0.072\pm 0.019$ &       $>1.33$      &     $<0.054$     \\ % 17 pix
    4       &      $42.3\pm 6.9$     &   $36\pm 5$    &  $\sim -247$   &    $2.35\pm 0.07$   &       $<38.2$         &       $<36.3$        &  $0.030\pm 0.006$ &       $>1.09$      &     $<0.027$     \\ % 32 pix
    5       &      $71.3\pm 8.7$     &   $56\pm 8$    &  $\sim -257$   &    $3.38\pm 0.11$   &       $<52.5$         &       $<64.3$        &  $0.035\pm 0.005$ &       $>1.34$      &     $<0.026$     \\ % 50 pix
    6       &      $75.6\pm 5.2$     &   $16\pm 1$    &  $\sim -224$   &    $1.97\pm 0.08$   &       $<13.3$         &       $<15.8$        &  $0.064\pm 0.007$ &       $>5.61$      &     $<0.011$     \\ % 45 pix
    7       &      $41.0\pm 4.8$     &   $23\pm 3$    &  $\sim -234$   &    $1.63\pm 0.07$   &       $<17.9$         &       $<14.9$        &  $0.042\pm 0.007$ &       $>2.26$      &     $<0.019$     \\ % 31 pix
    8       &     $194.1\pm 12.7$    &   $54\pm 3$    &  $\sim -291$   &   $11.93\pm 0.26$   &       $<89.4$         &       $<104.0$       &  $0.027\pm 0.002$ &       $>2.14$      &     $<0.013$     \\ % 144 pix
    9       &      $40.6\pm 6.6$     &   $47\pm 12$   &  $\sim -267$   &    $1.91\pm 0.06$   &       $<37.9$         &       $<41.0$        &  $0.036\pm 0.007$ &       $>1.06$      &     $<0.034$     \\ % 27 pix
    \hline                                                                                                                                  
    Total   &     $549.6\pm 21.5$    &                &                &   $25.36\pm 0.32$   &       $<314.2$        &       $<346.3$       &  $0.036\pm 0.002$ &       $>1.73$      &     $<0.021$     \\ \hline
    Average &                        &                &                &                     &                       &                      &  $0.043\pm 0.018$ &       $>1.51$      &     $<0.029$     \\ \hline
%    Total   &     $543.0\pm 25.5$    &   $92\pm 5$    &  $\sim -254$   &   $26.00\pm 0.45$   &       $<312.6$        &                      &  $0.035\pm 0.002$ &       $>1.72$      &     $<0.020$     \\ \hline
%    Stacked &     $573.2\pm 21.9$    &       -        &       -        &   $27.05\pm 0.31$   &       $<254.5$        &       $<260.4$       &  $0.035\pm 0.002$ &       $>2.22$      &     $<0.016$     \\ \hline
  \end{tabular}
  \tablefoot{
  The clump numbers are the same as those indicated in Fig.~\ref{moment0}. The integrated flux densities were derived by summing the channels where emission is detected, multiplied by the channel width. The velocity (relative to Centaurus A) and FWHM were estimated by fitting the spectrum with a single Gaussian profile. For the HCN(1--0) and SiO(2--1) emission, upper limits at $3\sigma$ were derived assuming the same FWHM and the same area as for HCO$^+$(1--0).
  The last three columns report the line ratios of the integrated intensities in $K.km.s^{-1}$ and are therefore corrected by the ratio of the rest frequencies (see Sect. \ref{sec:line_ratios}).
  }
\end{table*}

%%%---------------------------------------

\section{Morpho-kinematics of the gas}
\label{sec:Res}

   \subsection{HCO$^+$(1-0) emission}

   We first used the {\tt mapping} package from the GILDAS software\footnote{\url{http://www.iram.fr/IRAMFR/GILDAS}} to produce moment 0 maps of the HCN(1--0), HCO$^+$(1--0), and SiO(2--1) lines. We used the velocity range of the CO(1--0) emission in this region ($-350<v<-175\: km\,s^{-1}$; \citealt{SalomeQ_2017}) and a threshold of $1.1\sigma$. HCO$^+$ is detected, but HCN and SiO are not.
Using the moment 0 map of HCO$^+$(1--0) as a guide, we explored the data cube with the viewer tool in {\tt mapping} to constrain the velocity range of the line emission from each clump. For each clump, we then spectrally averaged the {\tt uv} table over the corresponding velocity range to produce an image of the emission.

   The signal-to-noise ratio of the data cube is rather low ($<$5 in each pixel). Therefore, we created a 3D mask to exclude channels that do not contain signal, to limit the impact of noise, and to produce more accurate moment maps of the HCO$^+$(1--0) emission. For each clump, the mask selects the channels within the velocity range of the line emission and the pixels within the $2\sigma$ contours of the spectrally integrated emission.

   Figure \ref{moment0} shows the moment 0 map of the CO(1--0) emission, along with the contours of the HCO$^+$(1--0) emission.
The HCO$^+$ emission is distributed into nine clumps and has the same morphology as the CO(1--0) emission, with the Horseshoe-like feature clearly identified.
Dense gas tracers have already been detected in molecular outflows (e.g. by \citealt{Salas_2014} in M82, by \citealt{Alatalo_2015b} and \citealt{Cicone_2020} in Mrk 231, by \citealt{Walter_2017} in NGC 253, and by \citealt{Barcos_2018} in Arp 220). However, this is the first detection of dense gas tracers in a region of jet-gas interaction outside the galactic plane.

   We extracted the integrated spectrum of each clump within the $2\sigma$ contour of this new moment 0 map. The spectra were analysed using the {\tt CLASS} package of GILDAS. The spectral resolution was first decreased to 6~km$\,s^{-1}$ in order to improve the signal-to-noise ratio without under-sampling the line. The integrated flux density $\int S_{HCO^+}d\varv$ was derived by integrating the spectra over the velocity ranges used to build the 3D mask. For the peak velocity $v_{peak}$ and the full width at half maximum (FWHM), we fitted the emission with a single Gaussian. The characteristics of the HCO$^+$(1--0) emission of each clump are reported in Table \ref{table:spec}.
The spectra are presented in Fig. \ref{clumps}.

   \subsection{HCN(1-0) emission}
   \label{sec:upper_limits}

   The HCN(1--0) line is not detected in these ALMA observations. Since the HCN and HCO$^+$ lines are commonly expected to trace dense gas, we assumed that the HCN(1--0) line would be emitted by the same region as the HCO$^+$(1--0), with similar linewidths (e.g. \citealt{Pety_2017,MJJD_2019}). We extracted the HCN(1--0) spectrum of each clump within the $2\sigma$ contour of the HCO$^+$(1--0) moment 0 map and derived an upper limit at $3\sigma$:
\begin{equation}
  \int S_{HCN}d\varv\,{\rm (mJy\,km\,s^{-1})}<\frac{\sqrt{2\pi}}{2.354}\times 3\,\sigma{\rm (mJy)}\,FWHM_{HCO^+}
\end{equation}
where we assume FWHM$_{HCN}$=FWHM$_{HCO^+}$ (i.e. the FWHM of the HCO$^+$ from Table \ref{table:spec}).
The HCN(1--0) integrated flux density is lower than $13-90\: mJy.km.s^{-1}$, similar to the upper limit of $33.5\: mJy.km.s^{-1}$ reported by \cite{SalomeQ_2016a} based on Australia Telescope Compact Array (ATCA) observations within the H\rmnum{1} shell.

   The SiO(2--1) line is not detected either. Using the same method as for HCN(1--0), we derived upper limits of $16-104\: mJy.km.s^{-1}$.

   \subsection{CO(1-0) emission}

   We smoothed the CO(1--0) cube to the spatial resolution of the HCO$^+$(1--0) and extracted the CO(1--0) spectra of the clumps within the $2\sigma$ contour of the HCO$^+$(1--0) moment 0 map. The CO emission covers the same range of velocities as the HCO$^+$ emission (see Fig. \ref{clumps}). The CO line profiles were fitted with a Gaussian profile. The integrated fluxes are reported in Table \ref{table:spec}. The integrated fluxes obtained when considering the ALMA data alone are 10\% to 30\% lower (not reported in Table \ref{table:spec}), likely due to the diffuse CO(1--0) emission.

\section{Line ratios}
\label{sec:line_ratios}

   We studied the ratios of the velocity-integrated intensities $I_{CO}$, $I_{HCO+}$, and $I_{HCN}$ (in $K.km.s^{-1}$) and compared them to literature values. The line ratios of the integrated intensities of the clumps can be derived from the integrated flux densities by taking the rest frequency of the lines into account:
%\begin{eqnarray}
%  I_{HCO+}/I_{CO}  &=& \frac{\int S_{HCO+}\,d\varv}{\int S_{CO}\,d\varv}\, \times\,  \cbra{\frac{\nu_{CO}}{\nu_{HCO+}}}^2 \\
%  I_{HCO+}/I_{HCN} &=& \frac{\int S_{HCO+}\,d\varv}{\int S_{HCN}\,d\varv}\, \times\,  \cbra{\frac{\nu_{HCN}}{\nu_{HCO+}}}^2 \\
%  I_{HCN}/I_{CO}   &=& \frac{\int S_{HCN}\,d\varv}{\int S_{CO}\,d\varv}\, \times\,  \cbra{\frac{\nu_{CO}}{\nu_{HCN}}}^2
%\end{eqnarray}
\begin{equation}
  I_1/I_2 = \cbra{\int S_1\,d\varv\; \bigg/ \int S_2\,d\varv}\, \times\,  \cbra{\nu_2/\nu_1}^2
,\end{equation}
where $I_i$ are the integrated intensities, $\int S_i\,d\varv$ are the integrated flux densities, and $\nu_i$ are the frequencies. The line ratios of each clump are reported in Table \ref{table:spec}.

\begin{figure*}[h]
  \centering
  \includegraphics[page=2,height=8cm,trim=10 15 30 40,clip=true]{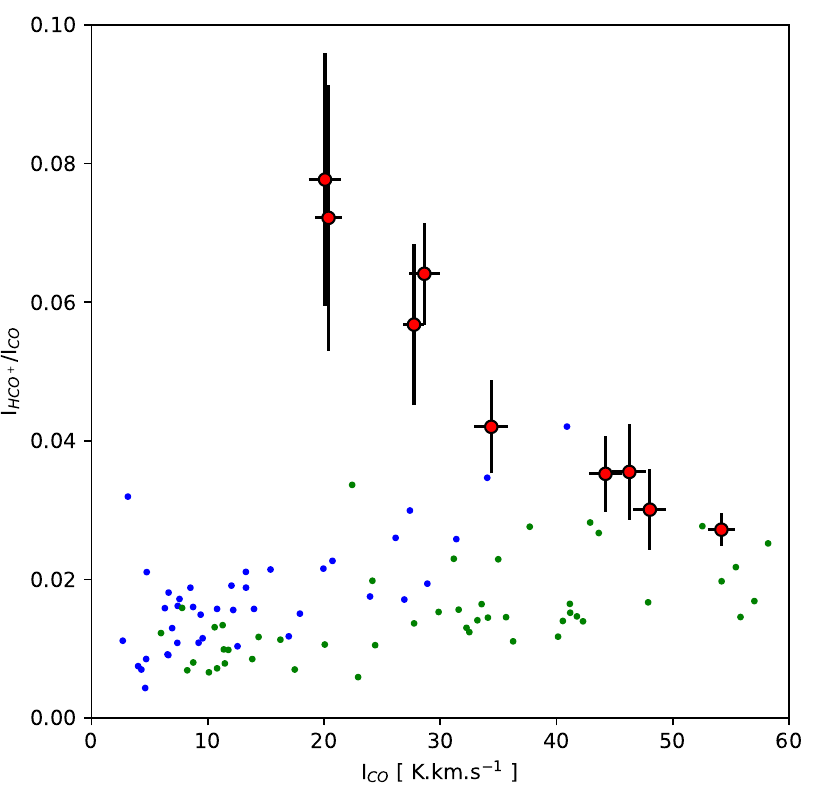}
  \hspace{3mm}
  \includegraphics[page=5,height=8cm,trim=10 15 35 40,clip=true]{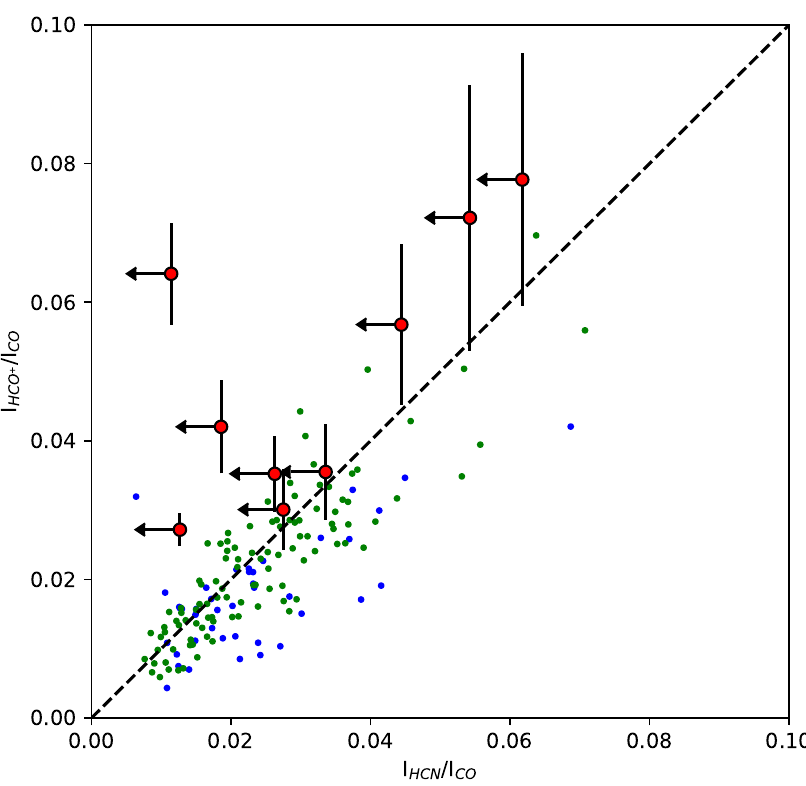}
  \caption{\label{relations} $I_{HCO+}/I_{CO}$ as a function of $I_{CO}$ (\emph{left}) and $I_{HCN}/I_{CO}$ (\emph{right}).
  The red points are the clumps in Table \ref{table:spec}, and the blue and green points come from the EMPIRE \citep{MJJD_2019} and ALMOND surveys \citep{Neumann_2023}. The dashed line in the right panel shows the unity relation.}
\end{figure*}

   The $I_{HCO+}/I_{CO}$ line ratio varies between 0.03 and 0.08. This is significantly higher than what was observed in nearby star-forming galaxies by the EMPIRE \citep{MJJD_2019} and ALMOND surveys \citep{Neumann_2023}. Moreover, $I_{HCO+}/I_{CO}$ decreases with increasing CO integrated intensity, contrary to what is observed in star-forming galaxies (Fig. \ref{relations} - left).
This suggests that the HCO$^+$(1--0) emission is not tracing the dense gas within the northern filaments of Centaurus A. Instead, it looks as if the HCO$^+$ emission is being enhanced by an external process that is not related to the density.

   For the $I_{HCN}/I_{CO}$ and $I_{HCO+}/I_{HCN}$ line ratios, we derived upper and lower limits, respectively. The line ratio $I_{HCN}/I_{CO}$ is lower than 0.06 and $I_{HCO+}/I_{HCN}$ is higher than unity. The average $I_{HCO+}/I_{HCN}$ of the clumps is $>1.51$, while the $I_{HCO+}/I_{HCN}$ ratio of the total emission from the clumps is $>1.72$. In particular, three clumps detected in HCO$^+$ have a line ratio $I_{HCO+}/I_{HCN}$ higher than 2, similar to that in the nuclear region of NGC\,5128 \citep{McCoy_2017}.
The $I_{HCO+}/I_{HCN}$ line ratio is significantly higher than the typical ratio observed in star-forming galaxies (Fig. \ref{relations} - right; \citealt{Brouillet_2005, MJJD_2019,Forbrich_2023,Garcia-Rodriguez_2023,Neumann_2023}).

   The $I_{HCO+}/I_{HCN}$ ratio for the total emission is typical of what is observed in starbursts \citep{Imanishi_2007,Salas_2014,Schirm_2016,Walter_2017} or 'composite' AGN associated with a nuclear starburst \citep{Kohno_2003,Kohno_2005,Krips_2008,Privon_2015}. Such high line ratios can also be associated with a low gas metallicity (e.g. \citealt{Galametz_2020,Forbrich_2023}). However, there is no evidence of recent star formation in the Horseshoe complex, and the gas metallicity is only slightly sub-solar (see \citealt{SalomeQ_2016b,SalomeQ_2017} and references within), suggesting that the enhanced HCO$^+$(1--0) emission likely has another origin.

The diffuse CO emission represents 10\% to 30\% of the total emission of the clumps. Therefore, if the HCN, HCO$^+$, and SiO emission is extended and not fully recovered by ALMA, our conclusion does not change since the $I_{HCO+}/I_{CO}$ ratio would be even higher.

\section{Low-velocity gradient modelling}
\label{sec:RADEX}

   We used the non-local thermal equilibrium radiative transfer code RADEX \citep{RADEX} to constrain the physical conditions of the clumps. Given a molecule, X, and the triplet \{$N_X$, $n_H$, $T_{kin}$\}, RADEX models the integrated intensities of the lines of the molecule. We adopted a linewidth $\Delta V=35\: km.s^{-1}$, which corresponds to the average FWHM of the clumps.
We compared the outputs of the RADEX models with the average integrated intensities over the clumps: $I_{CO}=36.0\pm 11.9\: K.km.s^{-1}$, $I_{HCO+}=1.56\pm 0.12\: K.km.s^{-1}$, and $I_{HCN}<1.03\: K.km.s^{-1}$.

   We explored the triplet \{$N_X$, $n_H$, $T_{kin}$\} with (i) $n_H=50,100,500,10^3$, and $10^4\: cm^{-3}$, (ii) $N_{CO}$ between $10^{15}$ and $10^{19}\: cm^{-2}$, and (iii) kinetic temperatures from $10\: K$ to $200\: K$.
We consider two phases of the gas that we call 'diffuse gas' ($n\lesssim 500\: cm^{-3}$; e.g. \citealt{Snow_2006}) and 'dense gas' ($n\geq 10^4\: cm^{-3}$; e.g. \citealt{Snow_2006}). This two phase are characterised by typical and significantly different relative abundances, N$_{HCO+}$/N$_{CO}$ and N$_{HCN}$/N$_{CO}$. We therefore derived the column density of HCO$^+$ and HCN from their relative abundances.
In the following, we adopt the relative abundances $N_{HCO+}/N_{CO}=10^{-3}$ and $N_{HCN}/N_{HCO+}=1.9$ \citep{Liszt_2010,Godard_2010} for the diffuse molecular gas and $N_{HCO+}/N_{CO}=3\times 10^{-5}$ and $N_{HCN}/N_{HCO+}=1.5$ \citep{Liszt_2010,Liu_2013} for the dense molecular gas.
When running RADEX, we took into account the excitation by collisions with H$_2$, He, and electrons. We adopted the following abundance ratios: $n_{He}/n_H=0.1$ and $n_e/n_H=0,10^{-5}$, and $10^{-4}$.

   The present grid of models only reproduces the observed average integrated intensities simultaneously for densities of $10^3$ and $10^4\: cm^{-3}$ (presented in Fig. \ref{radex}).
The models that reproduce the observations are highlighted in red and reported in Table \ref{table:radex_results}. Two regimes of the molecular gas can reproduce the observations: dense gas at $n_H=10^4\: cm^{-3}$ and $T_{kin}>65\: K$; and diffuse gas at $n_H=10^3\: cm^{-3}$ and $T_{kin}=15-165\: K$.
We get the same predictions regardless of whether the CO emission from ALMA+ACA or from ALMA alone is used.
We note that the predictions for the diffuse gas are in agreement with the mid-J CO line ratios at lower spatial resolutions from \cite{SalomeQ_2019}, which are indicated by the dotted contours in Fig. \ref{radex}.
\medskip

\begin{table}[h]
  \centering
  \caption{\label{table:radex_results} Predictions from RADEX.}
  \begin{tabular}{lcccc}
    \hline \hline
    $N_{HCO+}/N_{CO}$  &    $n_H$    &    $n_e$    &       $N_{CO}$        & $T_{kin}$ \\
                      & ($cm^{-3}$) & ($cm^{-3}$) & ($10^{16}\: cm^{-2}$) &    (K)    \\ \hline
    $3\times 10^{-5}$ &    $10^4$   &      0      &       8.5--9.6        &   $>104$  \\
    ("Dense" gas)     &    $10^4$   &     0.1     &       7.9--9.6        &   $>103$  \\
                      &    $10^4$   &      1      &       6.6--7.6        &   $>65$   \\ \hline
    $10^{-3}$         &    $10^3$   &      0      &       2.0--4.3        &  19--166  \\
    ("Diffuse" gas)   &    $10^3$   &     0.01    &       2.0--4.8        &  15--125  \\ \hline
  \end{tabular}
  \tablefoot{
  This table shows the triplets \{$N_{CO}$, $n_H$, $T_{kin}$\} that reproduce the CO and HCO$^+$ intensities and the upper limit on HCN for the diffuse and dense gas.
  }
\end{table}

   We explored the effects of small variations in relative abundance values around their typical estimates taken here to characterise the dense and diffuse gas. This is illustrated in Appendix \ref{sec:test_radex}.
For the dense gas, varying the abundance does not change the possible values of the gas density, which remain high ($\geq 10^4\: cm^{-3}$). However, if the abundance is significantly reduced (by 40\% or more), then there are no longer any compatible solutions at $n=10^4\: cm^{-3}$. On the other hand, if the abundance is increased significantly (by 40\% or more), then the possible temperature solutions tend to decrease and to become bounded.
For the diffuse gas, varying the abundance does not change the possible values of the gas density either: it remains low ($\leq 10^3\: cm^{-3}$). If the abundance is significantly reduced (by 40\% or more), then there are again no longer any compatible solutions, even at high densities. On the other hand, if the abundance is significantly increased (by 30\% or more), then solutions at lower temperatures and lower densities become possible.
In conclusion, abundances significantly lower than the typical values used for the dense and diffuse gas here are excluded or would lead to models not compatible with our observations. To the contrary, significantly higher abundances would imply possible lower temperatures for dense gas and lower temperatures and/or lower densities for diffuse gas. It should be noted, however, that these conclusions are based solely on variations around characteristic abundance values for two cases: dense and diffuse gas. A detailed study of the variation in abundance in a less constrained parameter space is not carried out here.
\medskip

\noindent \textit{SiO emission} - 
While SiO(2--1) is not detected, we derive upper limits (see Sect. \ref{sec:upper_limits} and Table \ref{table:spec}). We investigated the effect of considering an upper limit $I_{SiO}<1.08\: K.km.s^{-1}$ when constraining the RADEX. To do so, we used the SiO abundance $N_{SiO}/N_{H_2}=10^{-8.5}$ (from \citealt{Towner_2024}). The upper limit on SiO is indicated by the magenta line in Fig. \ref{radex}. This upper limit is not compatible with the predictions for the dense gas, suggesting that the detected HCO$^+$ emission is tracing diffuse molecular gas.
\medskip

\noindent \textit{Molecular gas mass} - 
Eight of the nine HCO$^+$ clumps seem to be unresolved with the present ALMA observations, for which the synthesised beam has a characteristic radius of 20 pc. The triplets \{$N_X$, $n_H$, $T_{kin}$\} from RADEX are average values of the total emission. Assuming spherical clouds, the average molecular gas mass of the clumps predicted by the models is $M_{pred}=8.3\times 10^5\: M_\odot$ for the diffuse gas abundances and $M_{pred}=8.3\times 10^6\: M_\odot$ for the dense gas abundances. Those estimates are larger than the average molecular gas mass derived from the CO emission with a standard CO-to-H$_2$ conversion factor: $M_{obs}=2.5\times 10^5\: M_\odot$. This suggests that these eight HCO$^+$ clumps are smaller than the beam of ALMA, and likely associated with diffuse gas.
\medskip

\noindent \textit{Total H$_2$ luminosity} - 
Cooling of the gas commonly occurs via H$_2$ emission. It is possible to estimate the total H$_2$ luminosity that a molecular cloud of radius $R=20\: pc$ would produce, assuming local thermal equilibrium:
\begin{equation}
  L_{H_2} = 4\pi\,D_L^2\Omega \cbra{\sum \frac{1}{4\pi}\,h\nu_{ij}\,A_{ij}\,N_i}
,\end{equation}
where $\nu_{ij}$ and $A_{ij}$ are the frequency and the Einstein A coefficient of the H$_2$ transition $i\rightarrow j$, $D_L$ is the luminosity distance, $\Omega$ is the solid angle of a clump of radius 20 pc, and $N_i$ is the column density of H$_2$ molecules in the excitation level $i$ given by
\begin{equation}
  N_i = N_{tot}\times \cbra{g_i\, exp\cbra{\frac{-E_i}{kT_{ex}}}}\; \bigg/\; \cbra{\sum g_i\, exp\cbra{\frac{-E_i}{kT_{ex}}}}
,\end{equation}
with $g_i$ and $E_i$ the weight and energy of level $i$, and $T_{ex}$ the excitation temperature. The total column density is estimated assuming a spherical cloud:
\begin{equation}
  N_{tot} = \frac{(4/3)\pi n R^3}{4\pi R^2}
.\end{equation}
\medskip

\noindent \textit{Cooling by H$_2$} - 
We predict that the total H$_2$ luminosity produced by a molecular cloud of radius 20 pc would be $L_{H_2}=2\times 10^{28}-3.7\times 10^{37}\: erg.s^{-1}$ if the gas is diffuse ($n_H=10^3\: cm^{-3}$ and $T_{kin}=T_{ex}=20-125\: K$), and $L_{H_2}=9.7\times 10^{37}-4.1\times 10^{39}\: erg.s^{-1}$ if the gas is dense ($n_H=10^4\: cm^{-3}$ and $T_{kin}=T_{ex}=100-200\: K$). We note that the H$_2$ luminosity is highly dependent on the gas temperature. Better constraints on the gas temperature are thus important, in particular for the diffuse gas solution. 
Observations with the K-band Multi Object Spectrograph (KMOS) of ro-vibrational lines of H$_2$ will allow the cooling energy radiated by H$_2$ to be measured (Salom\'e et al., in prep.). If the observed H$_2$ luminosity is lower than $10^{38}\: erg.s^{-1}$, we would be able to eliminate the dense gas solutions and constrain the gas temperature: $L_{H_2}=3.4\times 10^{30}\: erg.s^{-1}$ at 25 K, $7.1\times 10^{34}\: erg.s^{-1}$ at 50 K, $1.6\times 10^{36}\: erg.s^{-1}$ at 75 K, and $9.7\times 10^{36}\: erg.s^{-1}$ at 100 K.
\medskip

\noindent \textit{Heating by the radio jet} - 
The northern radio emission of Centaurus is complex, with several structures extended to different scales. \cite{Neff_2015a} estimated the total power within the different structures and reported a power $P_{NML}\sim 10^{44}\: erg.s^{-1}$ for the 'Northern Middle Lobe' over an area of $425\: kpc^2$.
If we consider a homogeneously distributed power within the Northern Middle Lobe, the H$_2$ emission produced by the dense gas would be at least 30\% of the power of the radio plasma available locally at the scale of the clumps. In particular, this fraction is higher than unity for a gas temperature higher than 120 K. This suggests that the energy of the Northern Middle Lobe is not enough to heat dense gas at $n_H=10^4\: cm^{-3}$.
Conversely, this fraction would be only 3\% for diffuse gas at 100 K. The energy injected by the radio jet is thus quantitatively a possible source of excitation of the H$_2$.
\medskip

\noindent \textit{Cosmic ray heating} - 
In the case of heating by cosmic rays \citep{Yusef-Zadeh_2007,Ferland_2008}, we can estimate the cosmic-ray ionisation rate, $\zeta$, needed to balance the cooling by the H$_2$ line emission.
We considered two cases representative of the diffuse and dense gas conditions predicted by RADEX: (i) $n_H=10^3\: cm^{-3}$ and $T_{kin}=T_{ex}=75\: K$, and (ii) $n_H=10^4\: cm^{-3}$ and $T_{kin}=T_{ex}=150\: K$.
Using the molecular gas mass and the total H$_2$ luminosity predicted by RADEX, the average total line emission per H$_2$ molecule would be $\mathcal{L}_{H_2}=3.2\times 10^{-34}\: W.H_2^{-1}$ for the diffuse gas and $\mathcal{L}_{H_2}=2.0\times 10^{-32}\: W.H_2^{-1}$ for the dense gas (Table \ref{table:H2_predictions}).

Following the discussion of \cite{Ogle_2010}, we estimated the cosmic ray ionisation rate using the following equation:
\begin{equation}
  \zeta = 1.2\times 10^{-13}\times \cbra{\frac{\mathcal{L}_{H_2}}{1.3\times 10^{-31}\: W}}\: s^{-1}.H^{-1}
.\end{equation}
To balance the H$_2$ cooling with cosmic ray heating, an ionisation rate $\zeta=3.0\times 10^{-16}\: s^{-1}.H^{-1}$ and $1.9\times 10^{-14}\: s^{-1}.H^{-1}$ is required for the diffuse and dense gas, respectively (Table \ref{table:H2_predictions}). This is respectively at least a factor of 0.4 and 24 greater than the cosmic ray ionisation rate derived in the centre of Centaurus A by \cite{vanderTak_2016}. This suggests that the heating of the molecular gas within the filaments is unlikely dominated by the cosmic rays.
\medskip

\begin{table}[h]
  \centering
  \caption{\label{table:H2_predictions} Cosmic ray heating predicted by RADEX, assuming local thermal equilibrium.}
  \begin{tabular}{lcccc}
    \hline \hline
       $n_H$    & $T_{ex}$ &     $L_{H_2}$       & $\mathcal{L}_{H_2}$  &       $\zeta$        \\
    ($cm^{-3}$) &   (K)    &   ($erg.s^{-1}$)    &    ($W.H_2^{-1}$)    &  ($s^{-1}.H^{-1}$)   \\ \hline
       $10^3$   &    75    & $1.6\times 10^{36}$ & $3.2\times 10^{-34}$ & $3.0\times 10^{-16}$ \\
       $10^4$   &   150    & $1.0\times 10^{39}$ & $2.3\times 10^{-32}$ & $1.9\times 10^{-14}$ \\ \hline
  \end{tabular}
  \tablefoot{
  We considered two sets of parameter characteristics for the diffuse and dense gas.
  }
\end{table}

\noindent \textit{Possible origin of the HCO$^+$ emission} - 
At densities $n\leq 10^4\: cm^{-3}$, high $I_{HCO+}/I_{HCN}$ line ratios can be explained by photo-dissociated regions (PDRs) or X-ray-dominated regions (XDRs; \citealt{Meijerink_2007}). In particular, the line ratios between dense gas tracers found in the nuclear region of NGC\,5128 by \cite{McCoy_2017} indicate that the HCO$^+$ might come from XDRs.
HCO$^+$ emission can also be enhanced by shocks in the diffuse gas \citep{Godard_2019}, and there is evidence of shocks in the filaments of Centaurus A \citep{Oosterloo_2005,SalomeQ_2016b,SalomeQ_2019}.
Studying the presence of PDRs, XDRs, or shocks in the northern filaments of Centaurus A is beyond the scope of the present paper and would need additional observational constraints. In particular, H$_2$ ro-vibrational lines can be used to identify the relative contribution of shocks and PDRs to the energy injection within the molecular gas \citep{Villa-Velez_2024}.

\section{Conclusion and discussion}
\label{sec:conclusion}

   We have presented the first observations of the HCN(1--0), HCO$^+$(1--0), and SiO(2--1) emission in the northern filaments of Centaurus A conducted with ALMA. HCO$^+$(1--0) is detected in nine clumps distributed along the Horseshoe complex, but no HCN or SiO is detected. We extracted the spectra of the clumps for the HCO$^+$(1--0) and CO(1--0) emission and computed upper limits of the HCN(1--0) and SiO(2--1) fluxes within the $2\sigma$ contours of the HCO$^+$ moment 0 map.
We derive relatively high $I_{HCO+}/I_{CO}$ and $I_{HCO+}/I_{HCN}$ line ratios and a relatively low $I_{HCN}/I_{CO}$ ratio compared to that typically observed in nearby star-forming galaxies. Moreover, we find that the $I_{HCO+}/I_{CO}$ ratio decreases with increasing $I_{CO}$. This indicates an enhanced HCO$^+$ emission that is likely not associated with a high dense-gas fraction.

   We used the average CO(1--0) and HCO$^+$(1--0) integrated intensities, as well as the upper limit on $I_{HCN}$, to constrain the grid of large velocity gradient models from RADEX. The observations can be explained either by diffuse molecular gas at $n_H=10^3\: cm^{-3}$ or by dense molecular gas at $n_H=10^4\: cm^{-3}$. However, we note that dense molecular gas is not compatible with the upper limit on $I_{SiO}$.
Moreover, an analysis of the predicted molecular gas mass and H$_2$ luminosity suggests that the HCO$^+$(1--0) likely arises from diffuse gas at $n\leq 10^3\: cm^{-3}$.

\begin{acknowledgements}
We thank the referee for his/her comments. \\

ALMA is a partnership of ESO (representing its member states), NSF (USA) and NINS (Japan), together with NRC (Canada), NSC and ASIAA (Taiwan), and KASI (Republic of Korea), in cooperation with the Republic of Chile. The Joint ALMA Observatory is operated by ESO, AUI/NRAO and NAOJ. \\

QS thanks the Observatoire de Paris for the access to the computational facilities. \\

Q.S. acknowledges the financial support from the visitor and mobility program of the Finnish Centre for Astronomy with ESO (FINCA), funded by the Academy of Finland grant nr 306531.
\end{acknowledgements}

\bibliography{Biblio}
\bibliographystyle{aa}

%\clearpage
\begin{appendix}
%\onecolumn

\section{HCO$^+$ clumps: Spectra}

\begin{figure*}[h!]
  \centering
  \includegraphics[height=3.7cm,trim=20 35 20 40,clip=true]{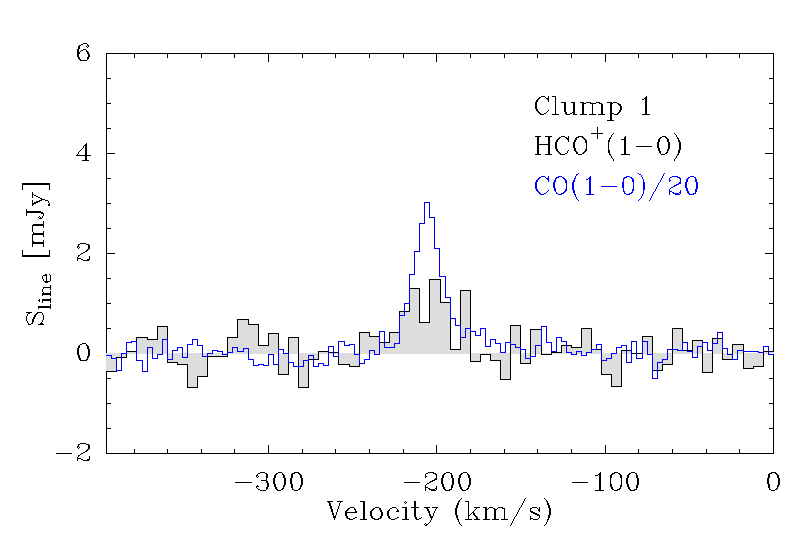}
  \hspace{2mm}
  \includegraphics[height=3.7cm,trim=20 35 20 40,clip=true]{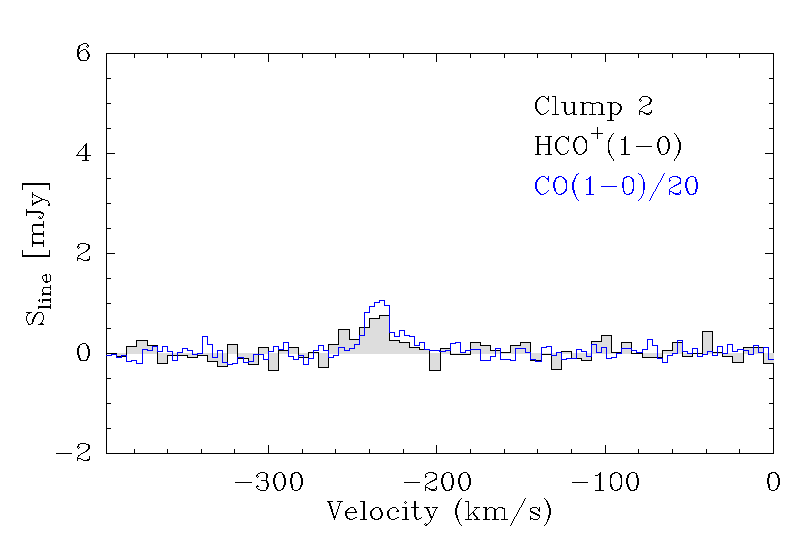}
  \hspace{2mm}
  \includegraphics[height=3.7cm,trim=20 35 20 40,clip=true]{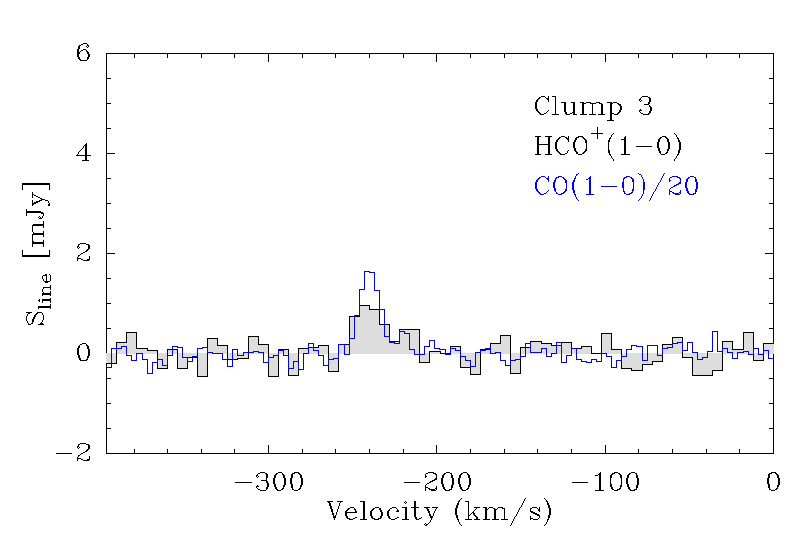} \\
  \vspace{3mm}
  \includegraphics[height=3.7cm,trim=20 35 15 40,clip=true]{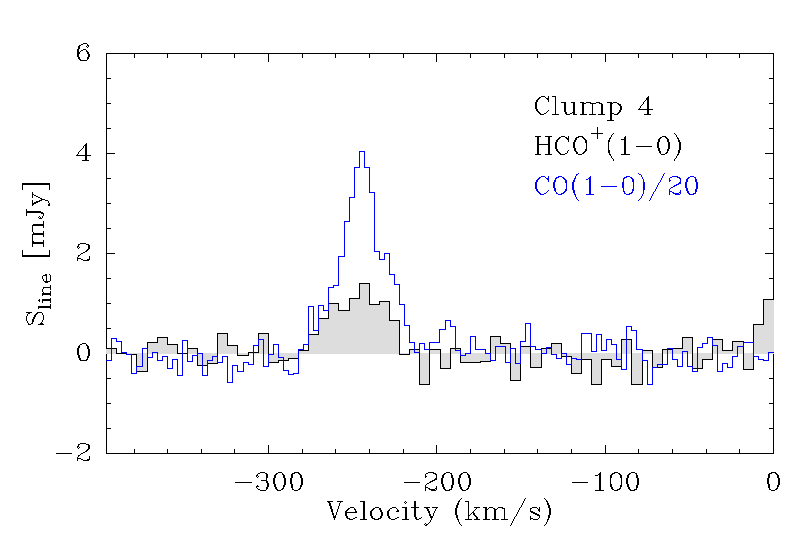}
  \hspace{2mm}
  \includegraphics[height=3.7cm,trim=20 35 15 40,clip=true]{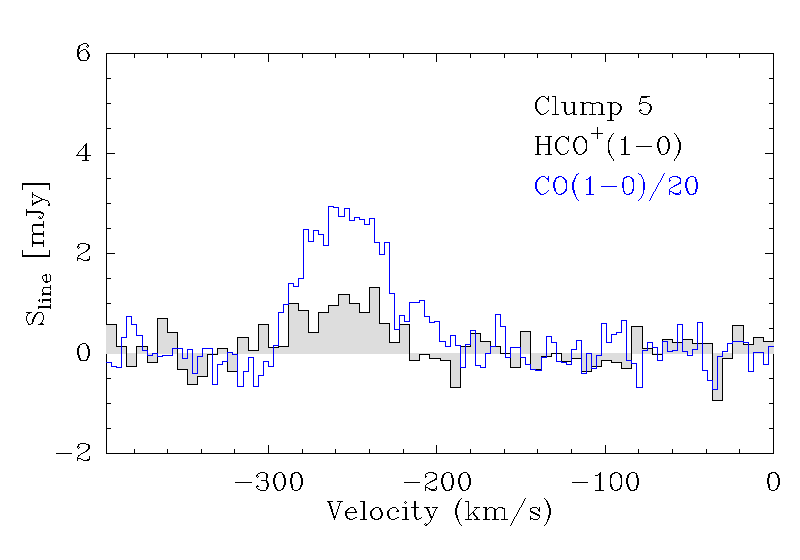}
  \hspace{2mm}
  \includegraphics[height=3.7cm,trim=20 35 15 40,clip=true]{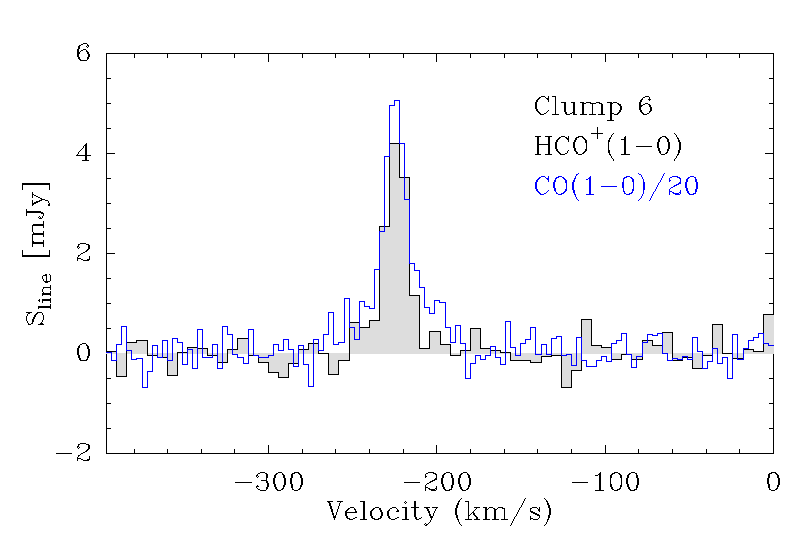} \\
  \vspace{3mm}
  \includegraphics[height=3.7cm,trim=20 35 20 40,clip=true]{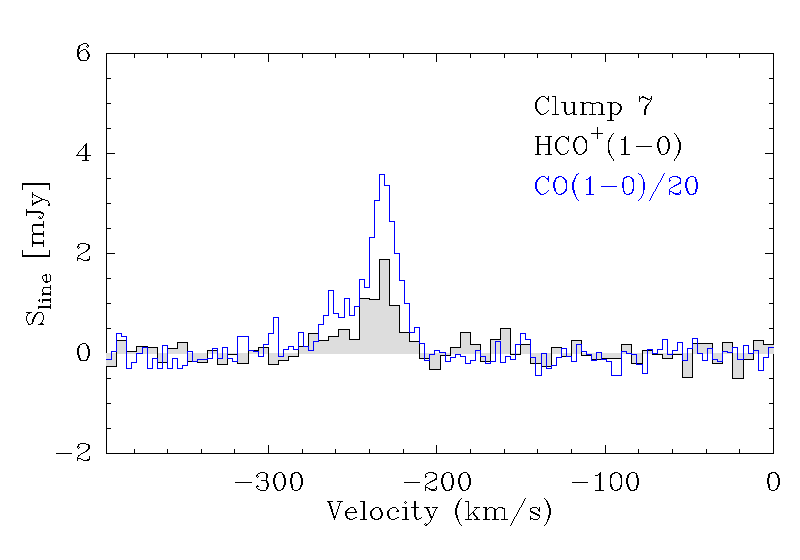}
  \hspace{2mm}
  \includegraphics[height=3.7cm,trim=20 35 20 40,clip=true]{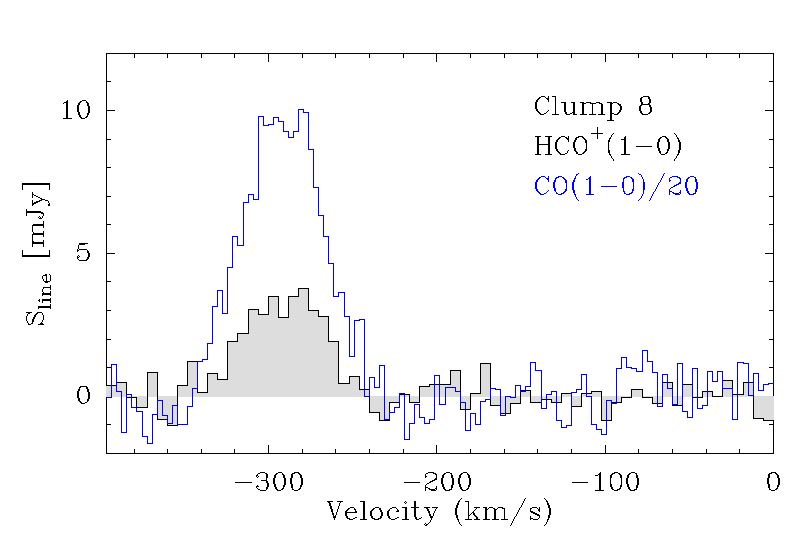}
  \hspace{2mm}
  \includegraphics[height=3.7cm,trim=20 35 20 40,clip=true]{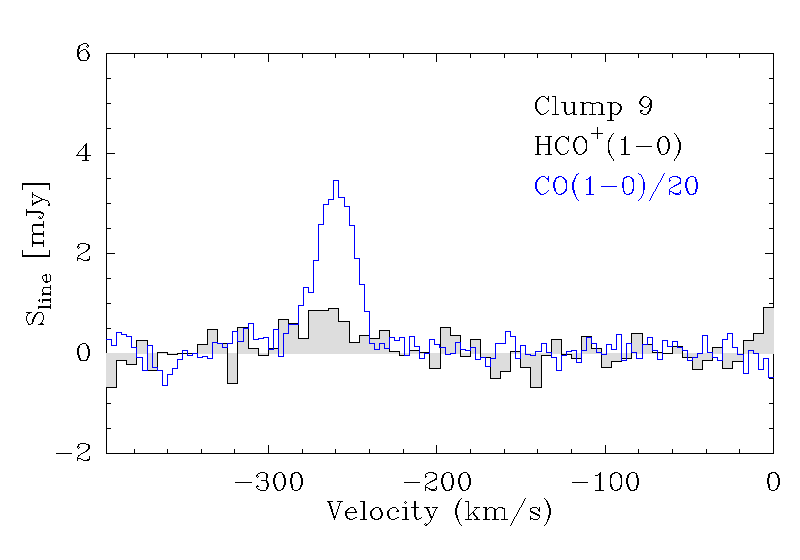}
  \caption{\label{clumps} Integrated spectra of HCO$^+$(1--0) and CO(1--0) are shown in black and blue, respectively. The CO(1--0) is scaled by a factor of 20. The velocities are relative to the systemic velocity of Centaurus A.}
\end{figure*}

\section{RADEX predictions}

\begin{figure*}[h!]
  \centering
  \includegraphics[page=5,height=7cm,trim=5 15 100 30,clip=true]{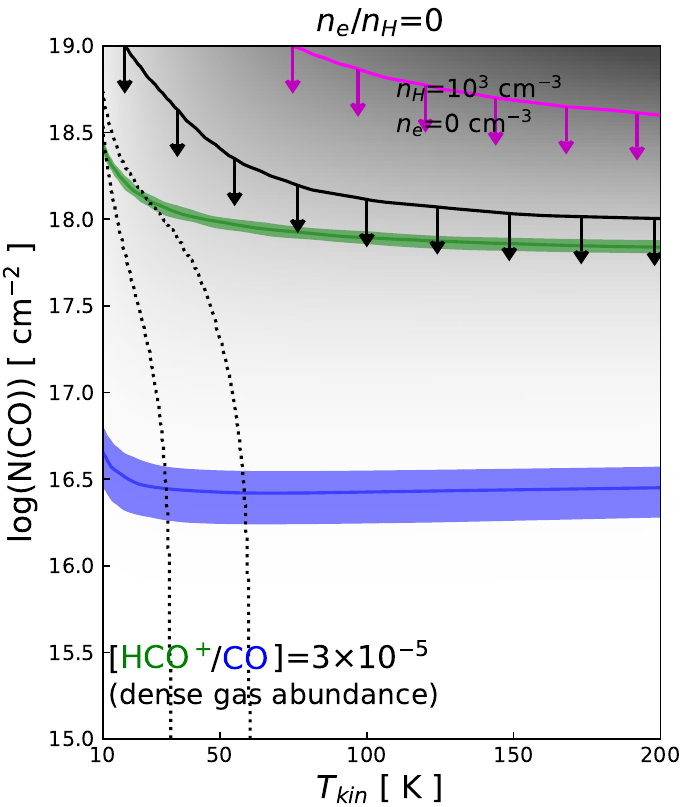}
  \includegraphics[page=5,height=7cm,trim=30 15 100 30,clip=true]{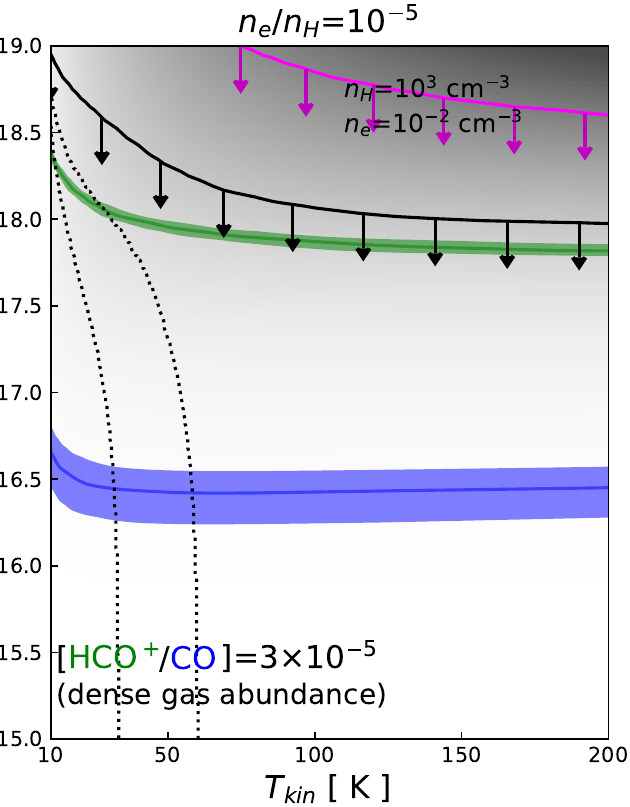}
  \includegraphics[page=5,height=7cm,trim=30 15 33 30,clip=true]{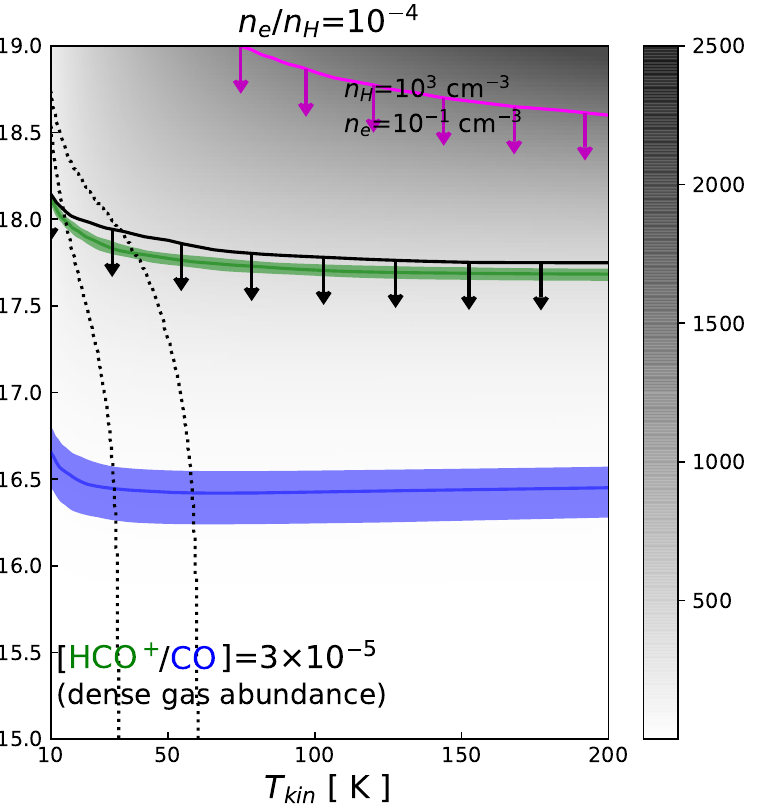} \\
  \includegraphics[page=17,height=7cm,trim=5 15 100 30,clip=true]{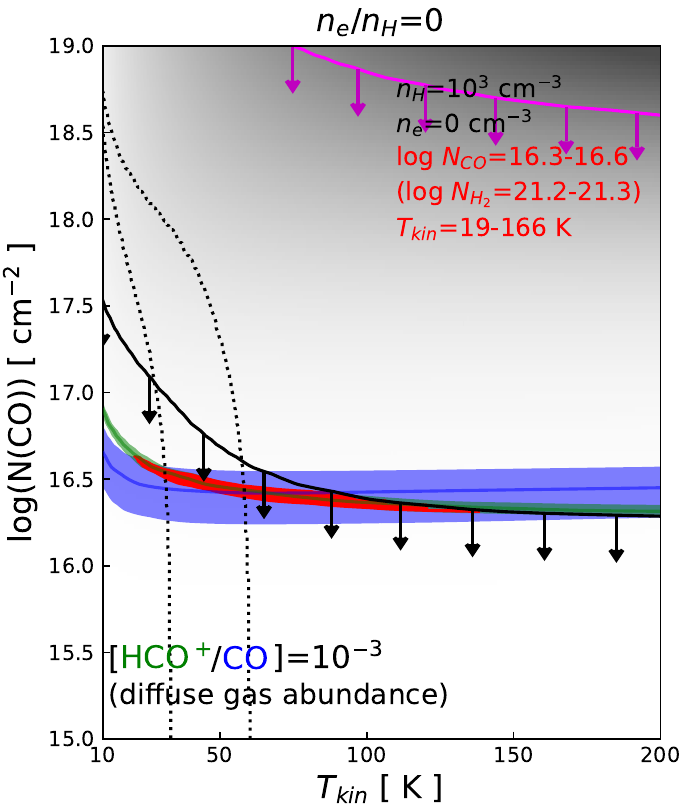}
  \includegraphics[page=17,height=7cm,trim=30 15 100 30,clip=true]{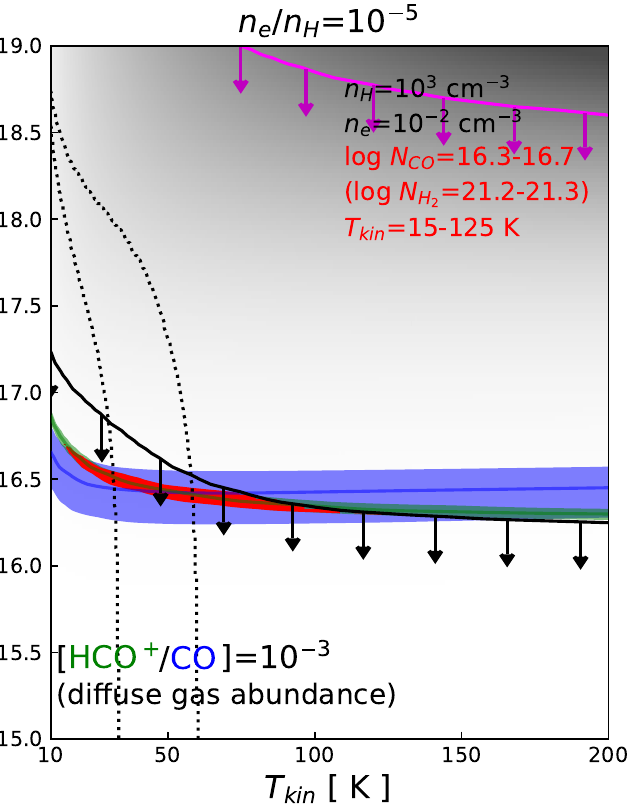}
  \includegraphics[page=17,height=7cm,trim=30 15 33 30,clip=true]{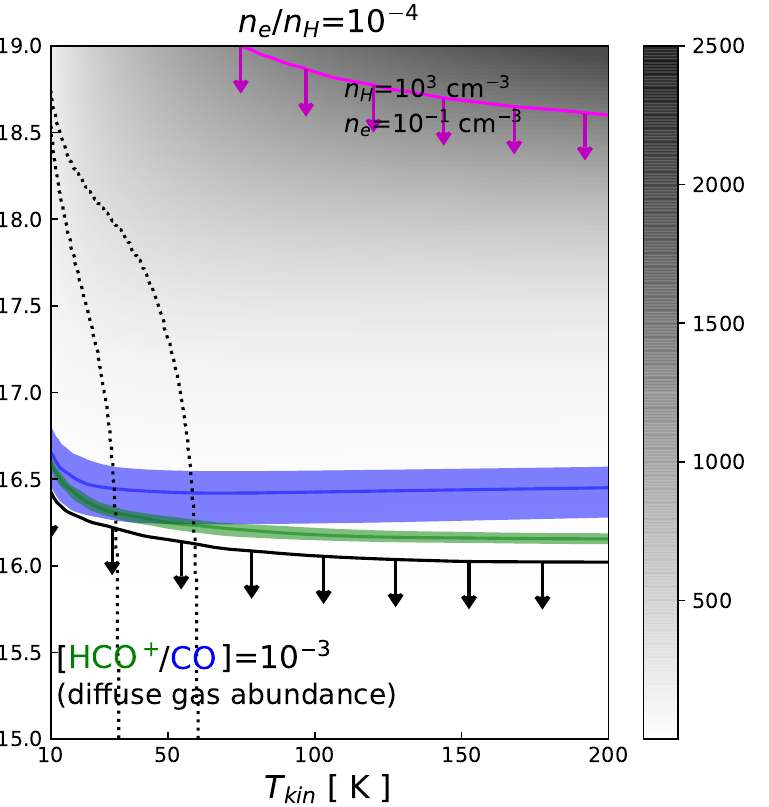}
  \caption{\label{radex} Radiative transfer predictions from RADEX for an HCO$^+$/CO abundance ratio typical of the dense (\emph{top}) and diffuse gas (\emph{bottom}).
  The blue and green areas show the range of observed CO(1--0) and HCO$^+$(1--0) integrated intensities, while the black and magenta lines indicate the upper limits of HCN(1--0) and SiO(2--1).
  The columns correspond to predictions for relative electron abundances of $n_e/n_H=0$ (left), $n_e/n_H=10^{-5}$ (middle), and $n_e/n_H=10^{-4}$ (right).
  The red areas correspond to the models that reproduce the CO and HCO$^+$ intensities and the HCN upper limit. The dotted lines indicate the models that reproduce the mid-J CO lines emission from \cite{SalomeQ_2019}.}
\end{figure*}

\begin{figure*}[h!]
  \centering
  \includegraphics[page=6,height=7cm,trim=5 15 100 30,clip=true]{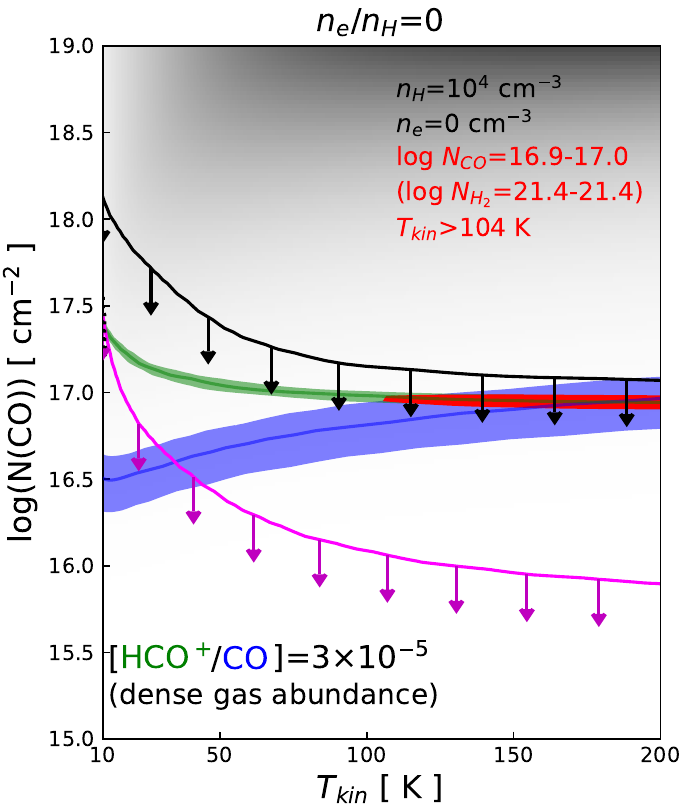}
  \includegraphics[page=6,height=7cm,trim=30 15 100 30,clip=true]{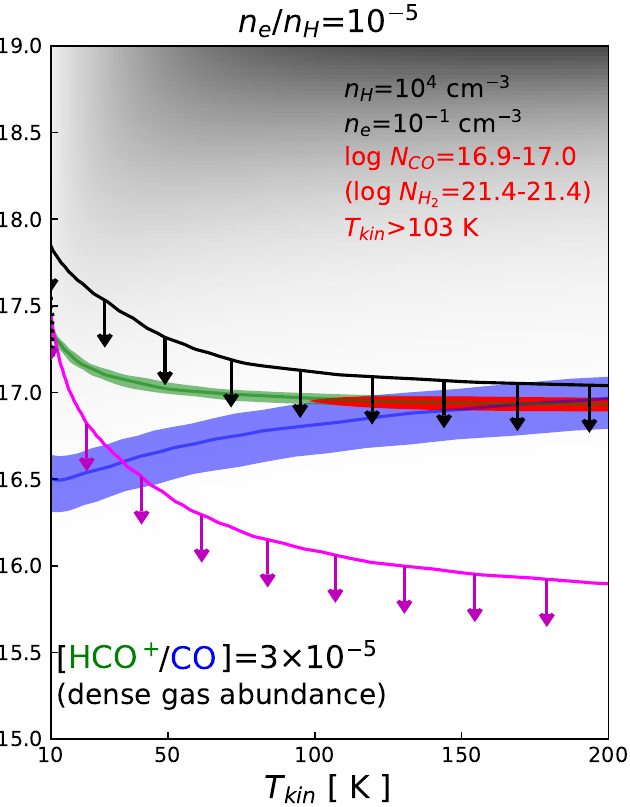}
  \includegraphics[page=6,height=7cm,trim=30 15 33 30,clip=true]{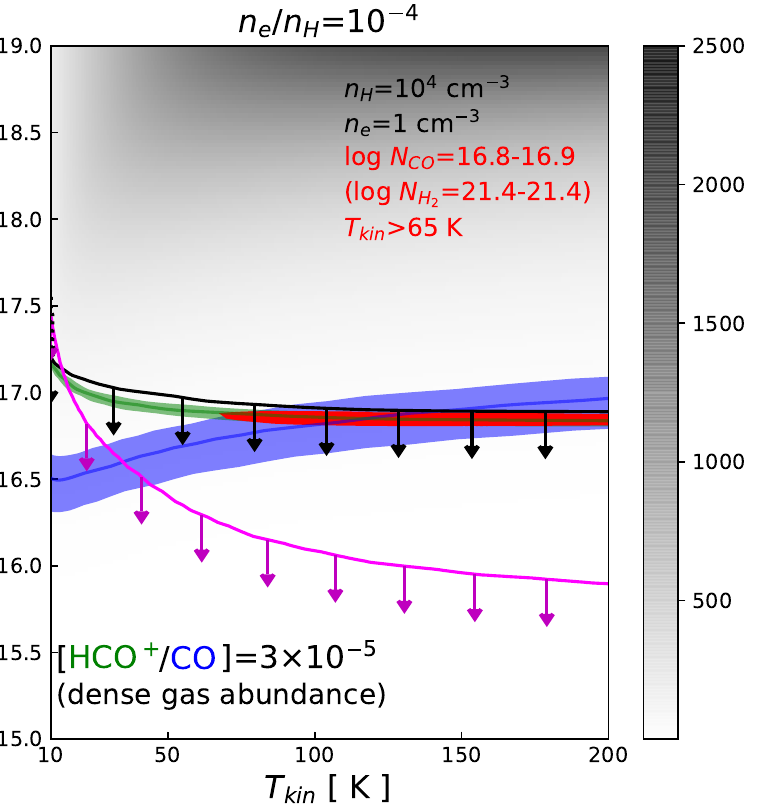} \\
  \includegraphics[page=18,height=7cm,trim=5 15 100 30,clip=true]{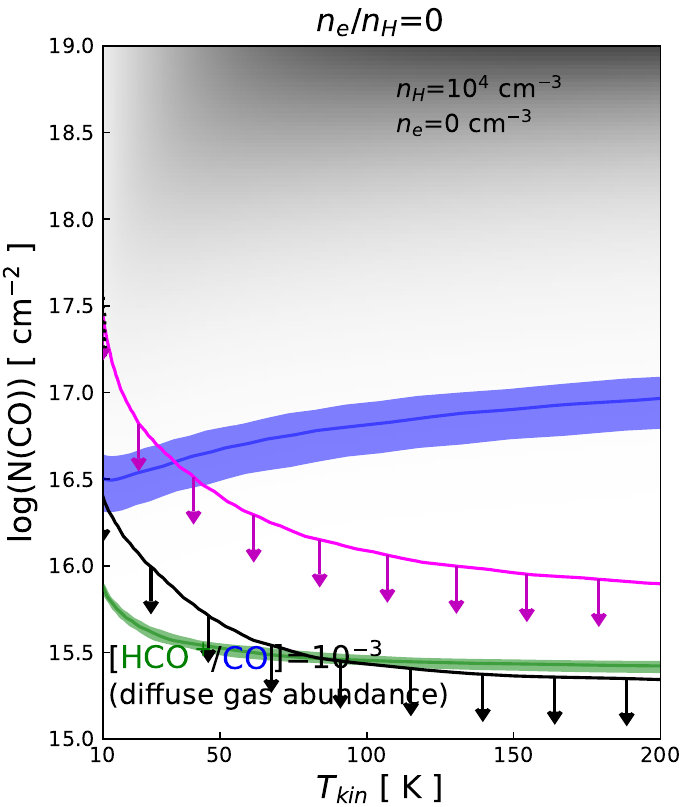}
  \includegraphics[page=18,height=7cm,trim=30 15 100 30,clip=true]{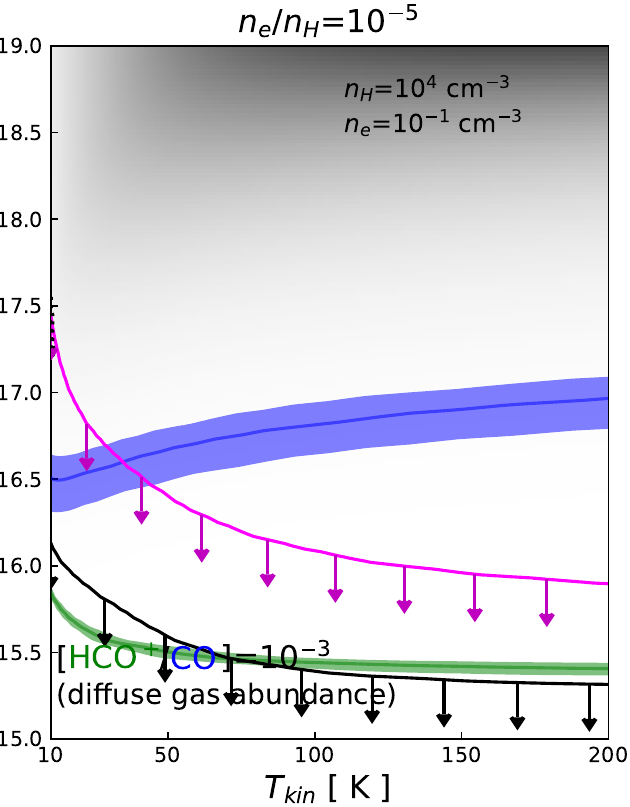}
  \includegraphics[page=18,height=7cm,trim=30 15 33 30,clip=true]{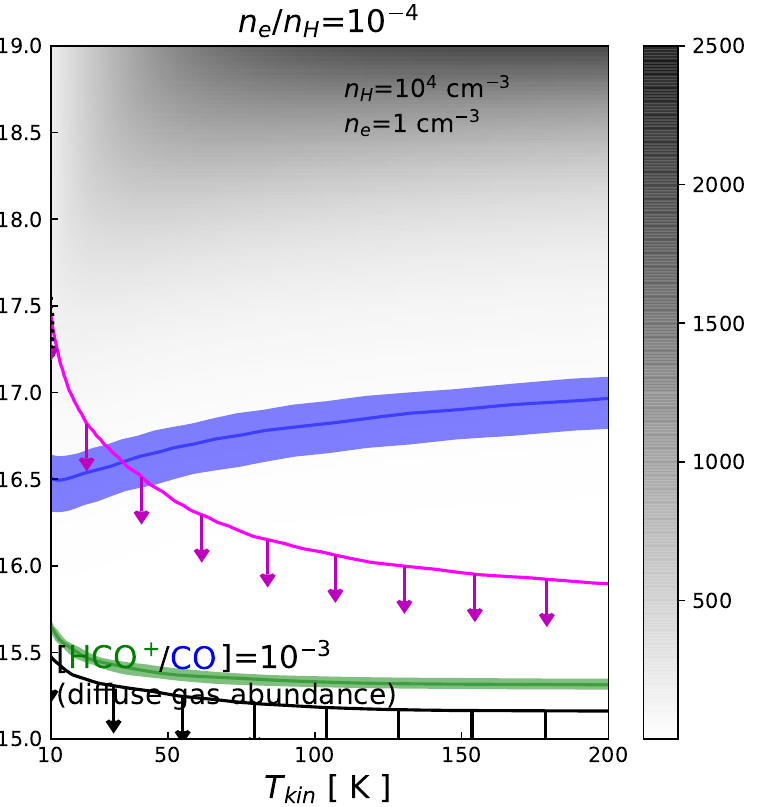} \\
  Fig.~\ref{radex}. Continued.
\end{figure*}

\section{Variation of the relative abundances}
\label{sec:test_radex}

\begin{figure*}[h!]
  \includegraphics[page=11,height=7cm,trim=5 15 100 30, clip=true]{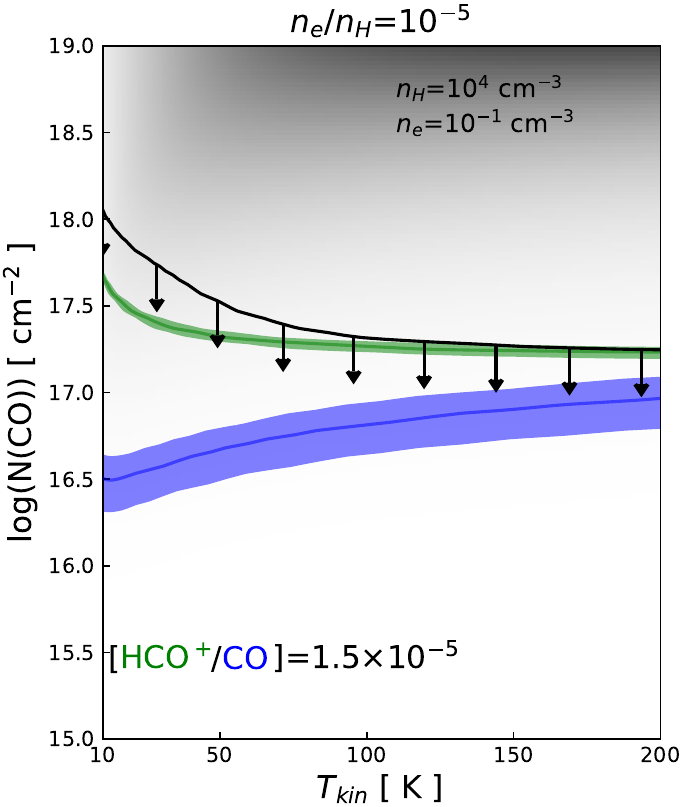}
  \includegraphics[page=12,height=7cm,trim=30 15 100 30,clip=true]{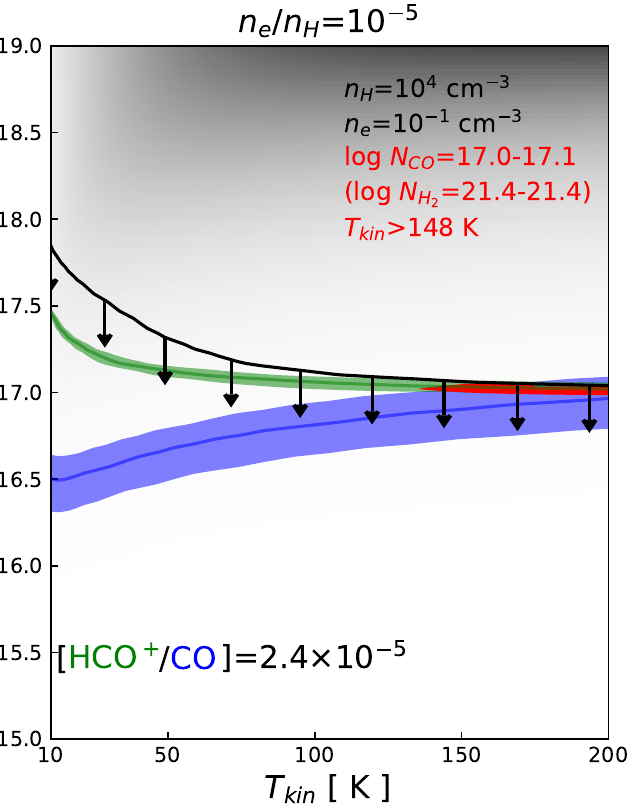}
  \includegraphics[page=13,height=7cm,trim=30 15 33 30, clip=true]{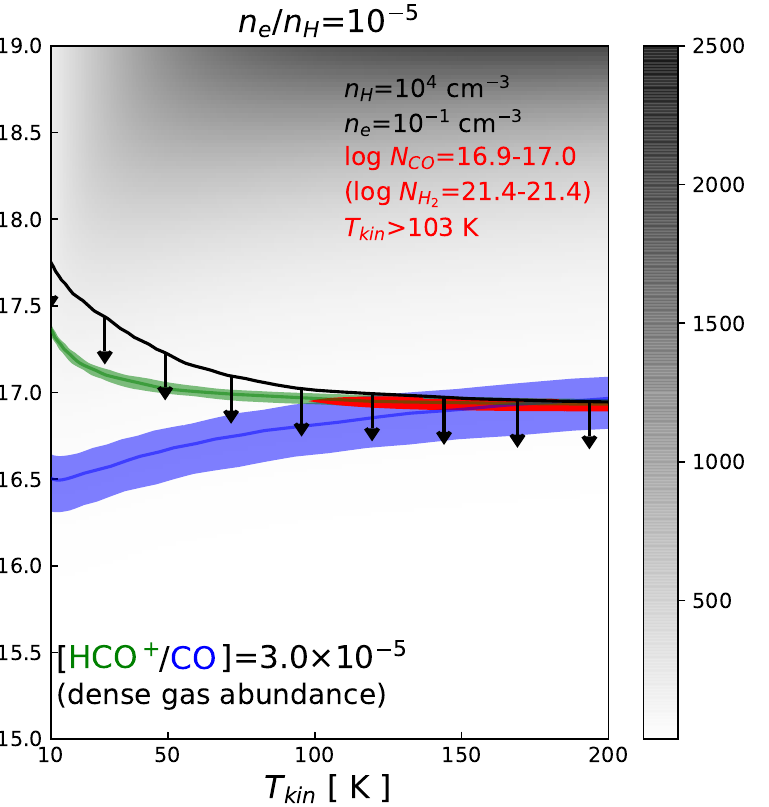} \\
  \includegraphics[page=14,height=7cm,trim=5 15 100 30, clip=true]{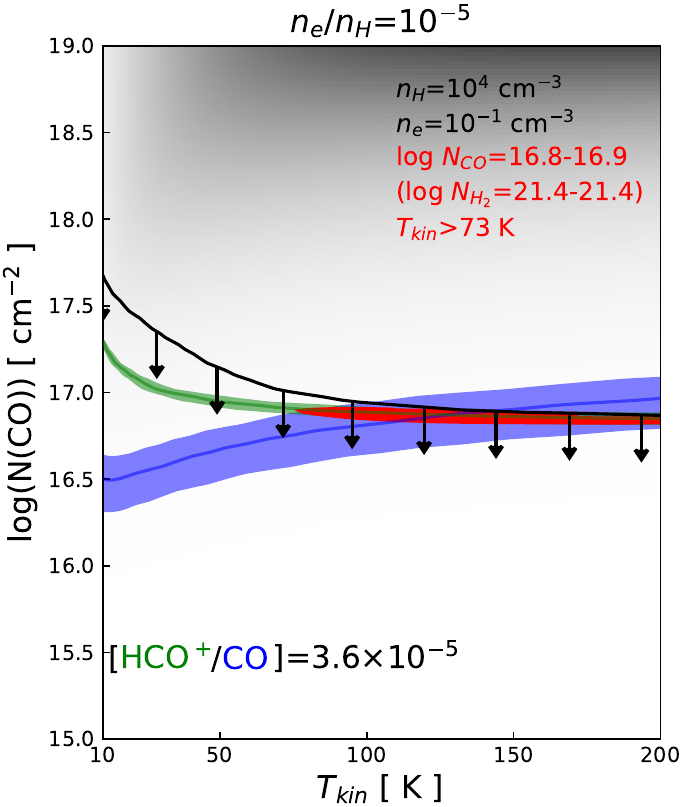}
  \includegraphics[page=15,height=7cm,trim=30 15 33 30, clip=true]{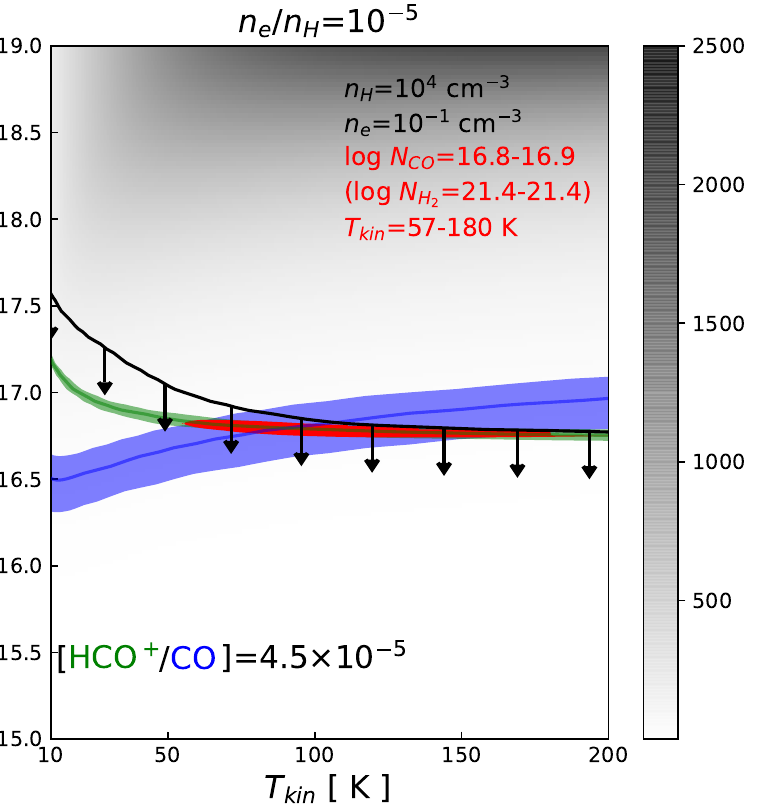}
  \caption{\label{test_radex} Radiative transfer predictions from RADEX for dense gas at $n_H=10^4$ with a relative electron abundance $n_e/n_H=10^{-5}$. The HCO$^+$/CO abundance ratio is respectively 50, 80, 100, 120, and 150\% of the typical value for the dense gas.
  The blue, green, and red areas and the black line are the same as in Fig. \ref{radex}.}
\end{figure*}

\begin{figure*}[h!]
  \includegraphics[page=6, height=7cm,trim=5 15 100 30, clip=true]{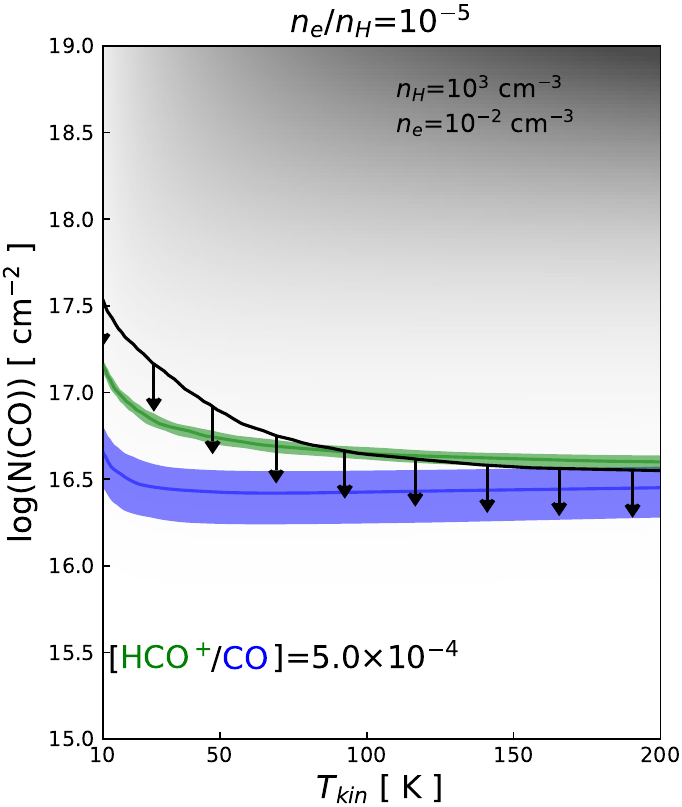}
  \includegraphics[page=7, height=7cm,trim=30 15 100 30,clip=true]{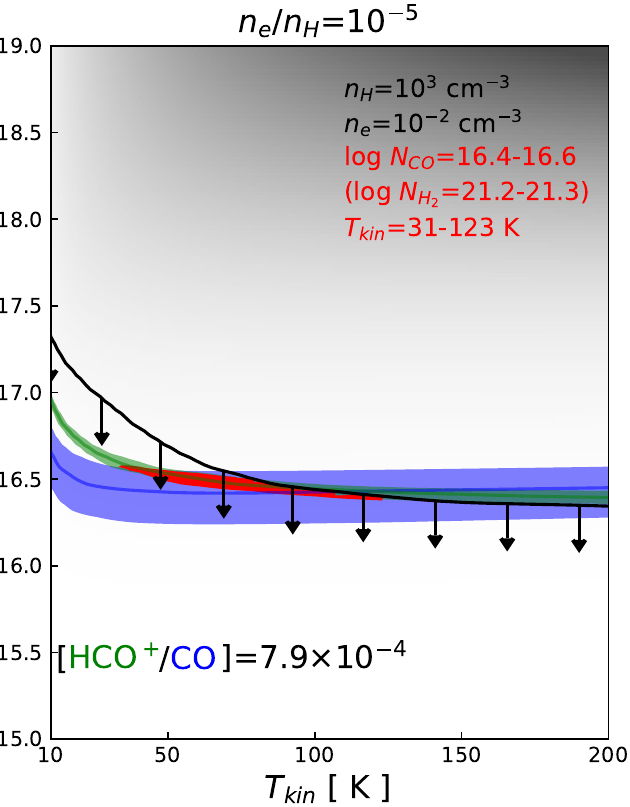}
  \includegraphics[page=8, height=7cm,trim=30 15 33 30, clip=true]{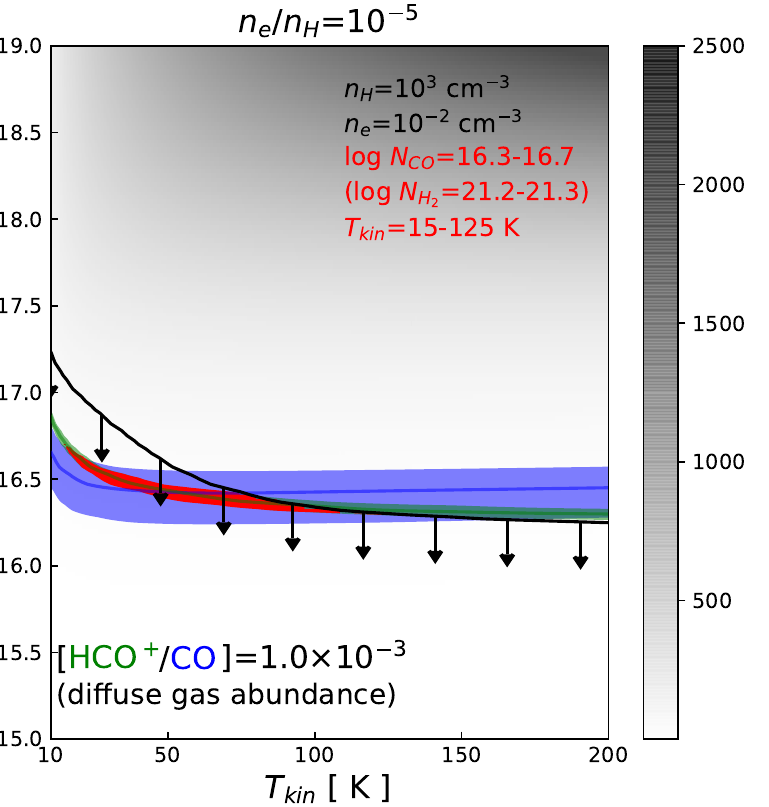} \\
  \includegraphics[page=9, height=7cm,trim=5 15 100 30, clip=true]{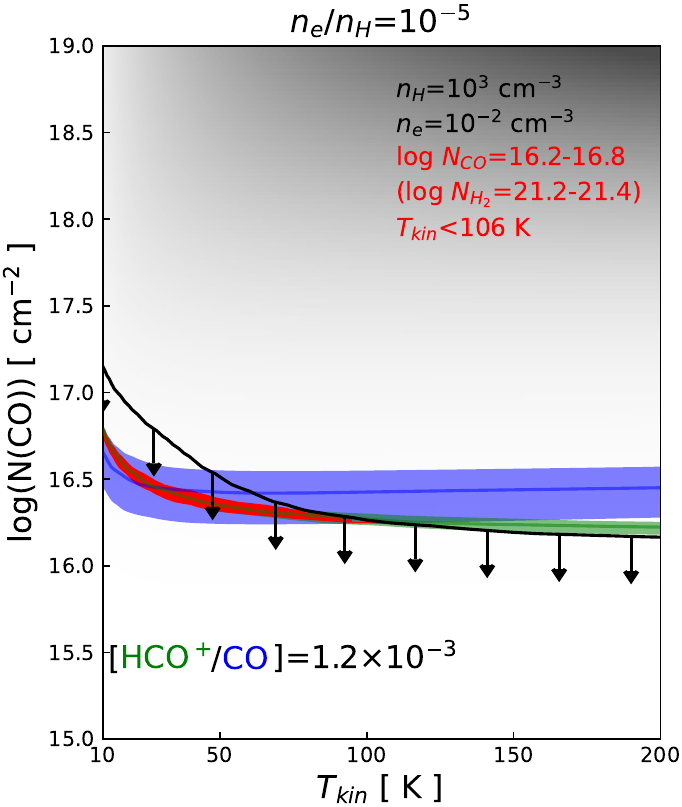}
  \includegraphics[page=10,height=7cm,trim=30 15 33 30, clip=true]{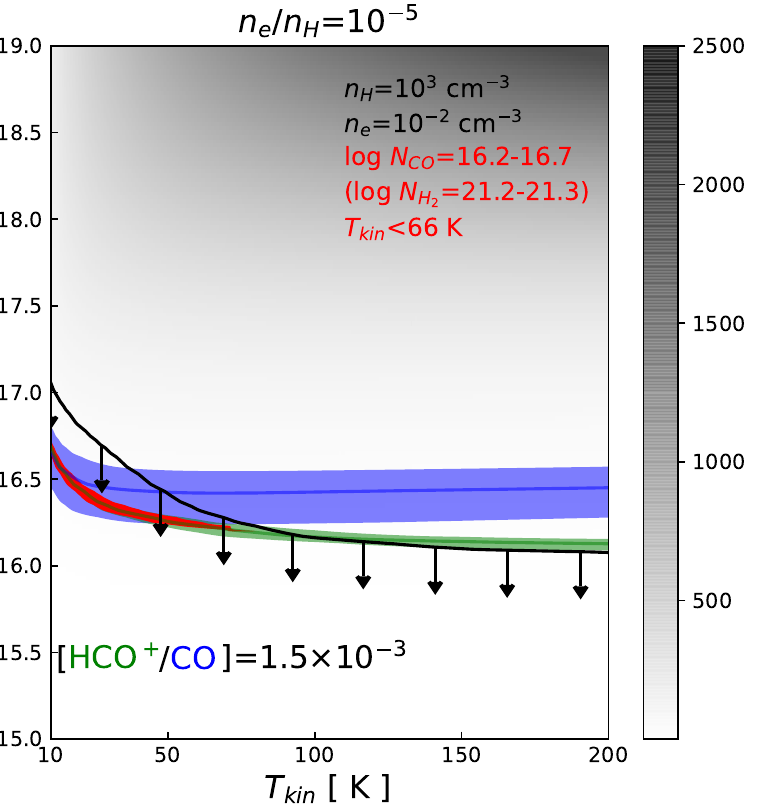}
  \caption{Same as Fig.~\ref{test_radex} but for diffuse gas at $n_H=10^3$.}
\end{figure*}

\end{appendix}
\end{document}